\documentclass[12pt,preprint]{aastex} 
\usepackage{hyperref}
 
\shorttitle{} 
\shortauthors{} 
 
\begin{document} 
 
\received{} 
\accepted{} 
 
\title{Tidally Excited Oscillations in Heartbeat Binary Stars: Pulsation Phases and Mode Identification }  
 
\author{Zhao Guo$^{1,2}$, Avi Shporer$^{3}$, Kelly Hambleton$^{4}$, Howard Isaacson$^{5}$}
\affil{${1}$ Center for Exoplanets and Habitable Worlds, Department of Astronomy \& Astrophysics, 525 Davey Laboratory, The Pennsylvania State University, University Park, PA 16802, USA \\
${2}$ Copernicus Astronomical Center, Polish Academy of Sciences, Bartycka 18, 00-716 Warsaw, Poland \\
${3}$ Department of Physics and Kavli Institute for Astrophysics and Space Research, Massachusetts Institute of Technology, Cambridge, MA 02139, USA \\
${4}$ Department of Astrophysics and Planetary Science, Villanova University, 800 East Lancaster Avenue, Villanova, PA 19085, USA \\
${5}$ Department of Astronomy, University of California, Berkeley CA 94720, USA
}
 
\slugcomment{11/18/2019} 

 
\begin{abstract} 
Tidal forces in eccentric binary stars known as heartbeat stars excite detectable oscillations that shed light on the processes of tidal synchronization and circularization.
We examine the pulsation phases of tidally excited oscillations (TEOs) in heartbeat binary systems. The target list includes four published heartbeat binaries and four additional systems observed by {\it Kepler}. To the first order, the pulsation phases of TEOs can be explained by the geometric effect of the dominant $l=2$, $m=0$, or $\pm 2$  modes assuming pulsations are adiabatic. We found that this simple theoretical interpretation can account for more than half of the systems on the list, assuming their spin and orbit axes are aligned. We do find significant deviations from the adiabatic predictions for some other systems, especially for the misaligned binary KIC 8164262. The deviations can potentially help to probe the non-adiabaticity of pulsation modes as well as resonances in the tidal forcing.

\end{abstract} 
 
 
 

\section{Introduction}

More than half of all stars reside in binaries, and tides can 
have a significant effect on stellar oscillations. In the first version of the classical textbook  {\it Nonradial oscillations of stars}, there is a whole chapter on tidal oscillations (Unno et al.\ 1979). However, it was removed in the second version (Unno et al.\ 1989), probably owing to the notion that such oscillations are difficult to be observed in practice. 

Tides can induce internal gravity waves (Zahn 1975; Goldreich \& Nicholson 1989; Goodman \& Dickson 1998; Ogilvie 2014). In early-type stars, these waves can form standing oscillation modes that can be observed at the stellar surface as tidally excited gravity modes. The dissipation of these gravity waves in the radiative envelope and near the surface causes the synchronization and circularization of the binary stars. Most studies on dynamical tides focus on the energy dissipation rate and the observed orbital parameter distribution of binaries and star-planet systems. The direct manifestation of equilibrium and dynamical tides can be shown in the light curves as flux variations. The theoretical foundations were laid out several decades ago (e.g., Press \& Teukolsky 1977), and Kumar et al.\ (1995, hereafter the Kumar LC model) even derived an expression for the observed flux variations from the tidal response. However, it is only after the {\it Kepler} satellite that we are able to observe unambiguously the tidally excited oscillations. The prototype system KOI-54 (Welsh et al. 2011), inspired lots of interests in observational (Hambleton et al. 2013, 2016, 2017) and theoretical studies (Fuller \& Lai 2012; Burkart et al. 2012; O'Leary \& Burkart 2014;  Fuller 2017;
Penoyre \& Stone 2018). Thompson et al.\ (2012) presented tens of such so-called `heartbeat' stars and the Kepler Eclipsing Binary (EB) Catalog (Slawson et al. 2011; Pr\v{s}a et al.\ (2011); Kirk et al. 2016) now consists of 173 such systems, flagged as `HBs'. Out of them, about 24 systems show tidally excited oscillations (TEOs)\footnote{Or called Tidally Induced Pulsations}, flagged as `TPs'. 

Most Kepler HBs are F- or A-type stars. Satellites surveying a larger portion of the sky have revealed massive heartbeat binaries of spectral type O and B. The eccentric binary with two O-type stars, $\iota$ Ori, has long been known to show periastron activities. The BRITE satellite observation revealed a detailed heartbeat feature in the light curve and tidally excited oscillations at about $20$ times of orbital frequency (Pablo et al.\ 2017). More massive HBs from BRITE are under study (Pigulski 2018). As for the all-sky survey of TESS, the first sector data already offered us a massive HB with TEOs (Jayasinghe et al.\ 2019).

While a lot of work has been focused on the frequencies and amplitudes of stellar oscillations, the pulsation phases also contain important information on the mode properties. When observed in multi-colour photometry, the phase difference in difference passbands, along with the amplitude ratios, can be used to identify the pulsation modes.
In binary stars, phase modulation can be used to detect the companions and derive orbital parameters. This has been applied to hundreds of binary stars as well as star-planet systems (Murphy et al.\ 2016, 2018). Bowman et al.\ (2016) and Zong et al.\ (2016) demonstrate that the amplitude and phase variations can be used to infer properties on the mode coupling, leading to the realm of non-linear asteroseismology. The ratio between the flux variation and the radial displacement on the stellar surface, i.e., the non-adiabatic parameter $f$, is a complex quantity, and its phase information has been used in the so-called `complex asteroseismic modeling' (Daszy{\'n}ska-Daszkiewicz et al.\ 2003; Daszy{\'n}ska-Daszkiewicz \& Walczak 2010). The phase changes due to eclipses in binary stars have the potential to facilitate the mode identification in the eclipsing systems (eclipse mapping, Biro \& Nusp 2011).

The HBs with TEOs are ideal laboratories to study the effect of equilibrium and dynamical tides. Unlike other stellar pulsations, the phases of TEOs can be predicted, and comparison with observations can be used to better understand mode physics. Only a few systems have been studied in detail, and the phase information is usually neglected. In this paper, we examine the pulsation phases of TEOs in eight HB systems and try to extract information on the mode identification.

\section{Analysis}  
\subsection{Equations}

Burkart et al.\ (2012) and O'Leary \& Burkart (2014) have derived the expressions for the pulsation amplitudes and phases of TEOs for adiabatic modes (see also Guo et al. 2017), and Fuller (2017) extended their work to the non-adiabatic modes and the cases of spin-orbit misalignment.
If we only consider the case when the pulsation, spin, and orbit axes are all aligned, the observed flux variations of TEOs at the stellar surface ($r=R$) can be expressed as: 
\begin{equation}
\frac{\Delta J(R)}{J(R)}=\left  [(2b_l-c_l)\frac{\xi_{r}(R)}{R} + +b_l \frac{\Delta F(R)}{F(R)}\right ] Y_{lm}(i_s,\phi_0)
\end{equation}

where $\xi_{r}(R)/R$ and $\Delta F(R)/F(R)$ are the relative Lagrangian radial displacement and the Lagrangian flux perturbation of the tidal response at the stellar surface, respectively. Note that the tidal response of the star, e.g., $\xi_{r}(r)$ and $\Delta F(r)$ for each forcing frequency, can be obtained from directly solving the non-adiabatic forced stellar oscillation equations (e.g., Pfahl et al.\ 2008; Vassechi et al.\ 2013) or from the mode decomposition method (Fuller 2017; Schenk et al.\ 2002). In the latter, the tidal response is expressed as a linear combination of free oscillation eigenfunctions, i.e., $\xi_{r}(r)=\sum_{\alpha} c_{\alpha}\xi_{r,\alpha}(r)$ (mode index $\alpha=(n, l, m)$). In the above equation, $b_l, c_l$ are the linear limb darkening coefficients (Burkart et al. 2012), and $i_s$ and $\phi_0$ are the orbital inclination and the observer's longitudinal coordinate, respectively.

The observed TEO phases arise from the two terms in eq.\ (1), 
\begin{equation}
arg\left (\frac{\Delta J(R)}{J(R)}\right )=arg \left[term1\right] + arg\left( Y_{lm}(i_s,\phi_0)\right)
\end{equation}

Under the assumption that: 

1) The pulsations are adiabatic and the observed TEOs are standing waves,

2) The observed TEOs are not fine-tuned (large detuning approximation), i.e., the difference between the driving frequency and the intrinsic eigenfrequency of the star is much larger than the mode damping rate,

then only the $Y_{lm}(i_s,\phi_0) \propto e^{im\phi_0}$ term contributes to the observed phases, since the mode adiabaticity and large detuning assumption imply $arg[term1]=0$ (term1 is real). When the flux variation is formulated as a sinusoidal function $a \sin [2 \pi (k f_{orb} t+\phi)]$, where $a$ and $\phi$ being the pulsation amplitude and phase, and $k f_{orb}$ being the pulsation frequencies of TEOs, the pulsation phases of TEOs for the dominant $l=2$ modes can be expressed as:

\begin{equation}
\phi_{l=2, m} =\left\{
\begin{array}{rcl}
0.25+m\phi_0  & & {\textrm{if}\  m=2 \ \ \textrm{or} -2}\\

0.25  & & {\textrm{if}\  m=0 }
\end{array} \right.
\end{equation}

where $\phi_0=0.25-\omega_p/(2\pi)$ and $m$ is the mode azimuthal number.

 All phases are measured with respect to the time of periastron passage $T_{peri}$ (implied when deriving Equation 1) and are in units of $2 \pi$. We cannot determine if the modes are retrograde ($m=2$) or prograde ($m=-2$) only from the phases \footnote{If we know the rotation rate of the star, we can usually determine if the modes are prograde or retrograde from the amplitude modeling of TEOs. From the theoretical and observational point of view, prograde modes are more likely.} as there is a $180^{\circ}$ phase ambiguity (Burkart et al.\ 2012). Thus a 0.5 phase offset ($180^{\circ}$) can be added or subtracted from equation (3) when comparing with observations. In the following, we use $m=2$ and $m=-2$ equivalently, unless we specify the mode is prograde or retrograde. 

Note that when assumption 1) is relaxed, we need to consider the radiative diffusion damping of gravity waves in the stellar interior, as well as the non-adiabatic effect near the stellar surface. This will introduce deviations from equation 3. The non-adiabatic effect will be addressed in the discussion section below.

By examining eq.\ (3), we see that accurate $T_{peri}$ and $\omega_p$ are needed to identify the pulsation modes from their phases. These two parameters can be obtained from radial velocity (RV) measurements as well as from the light curves (LCs). 

The dominant TEOs have a spherical degree of $l=2$, but which azimuthal number, $m$, do we expect? If the spin/pulsation and orbital axes are aligned, then only $m=0$ and $m=\pm 2$ modes are excited. And if we only consider the geometric effect (the $Y_{lm}(i_s,\phi_0) $ term), the observed mode amplitude is proportional to $\sqrt{\frac{(2l+1)(l-m)!}{4 \pi(l+m)!}} P^m_l(\cos i)$. When evaluated at $l=2$ and $m=0,2$, the amplitude ratio can be expressed as:

\begin{equation}
A_{m=0}/A_{m=2}= \frac{3 \cos^2 i -1}{\sqrt{\frac{3}{2}}\sin^2 i}
\end{equation}

Figure 1 shows the running of the amplitude ratio as a function of orbital inclination $i$. At low inclinations ($i< \approx 30$), $m=0$ modes are strongly favoured. But at intermediate to high inclinations, both $m=0$ and $m=2$ modes are expected, except for a limited range of $i \approx 51^{\circ} - 58^{\circ} $ when the $m=0$ modes reach the lowest amplitude contrast ($A_{m=0}/A_{m=2} < 5$).

\subsection{Methods}  
In order to identify the pulsation modes of TEOs, the parameters $T_{\rm peri}$ and $\omega_p$ are needed. Thompson et al.\ (2012) and Shporer et al.\ (2016) derived these quantities from the LCs and RVs, respectively. We chose among the HBs in the two papers that show obvious TEOs in the light curves. Since the TEOs are expected to be orbital harmonics, they should stand out even in the phase-folded light curves.   As we need to subtract the contribution of the equilibrium tide from the LC (usually modeled with the light curve synthesis code adopting the Roche model for the stellar shape, e.g., Wilson-Devinney (Wilson \& Devinney 1971), PHOEBE (Pr{\v s}a \& Zwitter 2005), and ELC (Orosz \& Hauschildt 2000)), we limit our samples to the HBs whose LCs can be fitted by the Kumar LC model reasonably well (unless detailed LC modeling was implemented). The Kumar model (Kumar et al.\ 1995) considers the tidal response of the star as a solution to a forced harmonic oscillator equation. Its final expression for the light curve takes into account the $l=2, m=0, \pm 2$ spherical harmonic components in the mode decomposition. Although mutual illumination (reflection effect) and eclipses are not taken into account in the model, we find that it is still a reasonable approximation for non-eclipsing and spin-orbit aligned systems. For systems with significant eclipses and mutual illumination, the aforementioned light curve tools (e.g., Wilson-Devinney) should be used. Our final list includes KIC 3230227, KIC 3749404, KIC 4248941, KIC 5034333, KIC 8719324, KOI-54, KIC 9016693, and KIC 8164262.  Among them, the TEO phases of KIC 3230227 and KOI-54 have been studied by Guo et al.\ (2017) and O'Leary \& Burkart (2014), respectively. Detailed binary modeling (LC+RV) have been performed for KIC 3749404 (Hambleton et al.\ 2016)  and KIC 8164262 (Hambleton et al.\ 2018; Fuller 2018). Additionally, we include the most massive HB binary $\iota$ Ori observed by the BRITE satellite (Pablo et al. 2017) and the earliest reported binary with TEOs, HD 209295 (Hander et al.\ 2002). We do not intend to perform a complete analysis of HBs with TEOs, but rather a first ensemble examination of the TEO phases in the most TEO-dominated HBs.
 
We de-trended the {\it Kepler} light curves (LCs) from Quarter 1 to Quarter 17, following the treatment in Guo et al.\ (2016, 2017).  The de-trended LCs were fitted with the Kumar light curve model (Kumar  et al.\ 1995, eq. 44; Thompson et al.\ 2012, eq. 1) 
to derive the estimated $T_{peri} , e, \omega_p$, and the orbital inclination $i$\footnote{Ideally, detailed binary modeling (a joint LC+RV fit) with the binary synthesis code including the tidal deformation and the reflection effect is needed to derive these parameters accurately. However, this is highly non-trivial, and we instead use the simple Kumar model to derive estimated values of these parameters.  Comparing to the parameters derived from RVs in literature, we find that the $e$ and $\omega_p$ from the Kumar model are usually reasonable.} These parameters can also be derived from the radial velocities (RVs) except for $i$. The orbital periods are adopted as the values in the Kepler EB catalog. We use the literature values for the above parameters if available. We specify the source as LC or (LC+RV) in Table 1-8. The uncertainties of $T_{peri}$ and $\omega_p$ are propagated to those of the theoretical TEO phases $\phi_m$. The binary light curve model (the part due to equilibrium tide) was subtracted from the de-trended light curves, and the Fourier spectrum of the residuals was calculated with the {\it Period04} package (Lenz \& Berger 2005). A standard pre-whitening procedure was performed to derive the pulsation frequencies, amplitudes, and phases. We fit the background noise of the Fourier spectrum in the log-log space using a Lorentzian-like function as in Pablo et al.\ (2017), Bowman et al.\ (2019), and Hander et al.\ (2019). The uncertainties of frequencies, amplitudes, and phases are estimated following Kallinger et al.\ (2008). The theoretical pulsation phases $\phi_m$ for $m=0, 2$ and $-2$ are calculated, plotted in Figures 3, 5, 7, 9, and $10-13$ as vertical strips, and listed in Table $1-8$. The width of the strips signifies the one sigma errors of theoretical phases. Also note that there are two possible solutions for the TEO phases with a difference of 0.5.

\section{Results}

\subsection{Four {\it Kepler} Heartbeat Binaries with TEOs}

\textbf{KIC\ 8719324} (Figure 2, 3, Table 1)

KIC 8719324 is a short-period binary with $P \approx 10$ days. A preliminary examination of the spectra from Keck HIRES shows that this binary consists of an F-star primary with an M-star companion (Guo et al.\ in prep.).
Figure 2 shows the phase-folded {\it Kepler }light curve. The dip and bump have a similar amplitude which signatures a system with an intermediate-to-high inclination angle. Actually, there is a grazing eclipse in the light curve at phase $\phi=0.95$. A fit with the Kumar light curve model yields an inclination of $73^{\circ}$. The orbital parameters are summarized in Table 1.

Strong oscillations are present, even in the phased folded light curve, indicating frequencies are of integer times of orbital frequency.  The Fourier spectrum in the bottom panel of Figure 2 shows two dominating TEOs at $N=26$ and $N = 29$ orbital harmonics. The pulsation phases of these two TEOs shown in Figure 3 suggest that they are likely $m=0$ and $m=2$ modes, respectively. This is in line with an intermediate-to-high inclination, for which both $m=0$ and $|m|=2$ modes are expected, but $|m|=2$ modes are preferred (larger amplitude). The two frequency peaks at $f < 1$ day$^{-1}$ ($f_1=0.323$ day$^{-1}$, $f_2=0.646$ day$^{-1}$) are likely due to rotational modulations (Zimmerman et al.\ 2017).

\textbf{KIC\ 9016693} (Figure 4, 5, Table 2) 

Figure 4 shows the light curve of KIC 9016693. It is a 26-day binary system, with an eccentricity of about $0.7$. The overall shape of the LC is a near-symmetric periastron brightening, indicating that it is nearly a face-on system. We obtain an inclination of 25 degrees from the Kumar model fit to the LC. 

A very strong pulsation in the LC reaches an amplitude of 0.2 milli-mag, and this strong sinusoidal shape of pulsation distorts the underlying binary light curve (the brightening feature).  
This dominating pulsation has a frequency of $f=0.91$ day$^{-1}$, which is exactly 24 times of orbital frequency $f_{orb}=0.03792$ day$^{-1}$. The pulsation phase is 0.275, which is very close to the prediction for an $m=0$ mode ($\phi=0.25$), and this complies with the low inclination angle of the binary (Figure 5).

\textbf{KIC 4248941}  (Figure 6, 7, Table 3)

The light curve in Figure 6 shows TEOs with very large amplitudes. We cannot remove the binary light curve perfectly, and this is why there are many orbital harmonics in the Fourier spectrum of the residuals. However, the $N = 5$ orbital harmonic stands out clearly and cannot be explained by the imperfect removal: it has to be a TEO. Figure 7 shows that its pulsation phase agrees with an $m=2$ mode.

\textbf{KIC 5034333}  (Figure 8, 9, Table 4)

The derived argument of periastron ($\omega_p$) of this binary is $278^{\circ}$, and thus the predicted TEO phases of $l = 2, m = 0$ and $m=\pm2$ modes are very close. The observed TEO phases shown in Figure 9 do tend to be close to the $m = 0$ and $2$ strips, although the $N=66, 19, 20$ TEOs deviate significantly from the predictions. Previously, Zimmerman et al.\ (2017) identified the signature of rotational modulations in the Fourier spectrum and measured the rotation period of the two components to be $3.98$ and $15.2$ days.

\subsection{Four Published Heartbeat Stars with TEOs}

The face-on system, KOI-54, has been discussed thoroughly in Fuller \& Lai (2012), Burkart et al.\ (2012), and O'Leary \& Burkart (2014). The two dominant TEOs have been identified as $l=2, m=0$ modes. The edge-on system KIC 3230227, presented in Guo et al. (2017), shows more than ten TEOs. Most of them are orbital harmonics and can be explained by $l=2, m=2$ prograde modes. Recently, TEOs in the $\delta$ Sct/$\gamma$ Dor hybrid pulsating eclipsing binary KIC\ 4142768 have also been identified as $l=2, m=2$ prograde modes (Guo et al.\ 2019).

Except for the above systems, a few other heartbeat binary stars have been characterized in detail. But no discussions on the pulsation phases are presented. Pablo et al.\ (2017) modeled the pulsation amplitudes of the TEOs in $\iota$ Ori and found that they agree with $l=2,m=2$ modes.  In this section, we present the pulsation phase analysis of TEOs in four published heartbeat binary systems.  

\textbf{HD 209295}: (Figure 10, Table 5)

This 3-day, eccentric ($e=0.35$) binary consists of an A-type $\gamma$ Dor/$\delta$ Scuti hybrid pulsating primary and probably a white dwarf companion (Handler et al.\ 2002). Although there are no 'heartbeat' features shown in the light curve (Handler, private communication), this binary was observed in multiple colours (B, V, and I-band). The Fourier spectrum shows five significant TEOs ranging from 3 to 9 times of orbital frequency.

We show the pulsation phases in three passbands in Figure 10. Although the $N= 3, 5, 9$ orbital harmonic TEOs agree with $|m|=2$, the one with the largest amplitude ($N=9$) has a phase close to that of an $m=0$ mode. Handler et al.\ (2002) only constrained the inclination as $i <  (40^{\circ}-45^{\circ})$. At this intermediate inclination, both $m=0$ and $m=2$ modes can reach observable amplitudes. The theoretical modeling of the tidally excited radial velocity amplitude (Willems \& Aerts 2002, Section 5.2) indicates that TEOs at $N=3, 4$ times of orbital frequency may be accounted for by $l=2, m=-2$ modes. This is in agreement with our phase measurements here. The observed pulsation phases do not seem to depend on the passbands, in line with the theoretical interpretation that it is mostly a geometric effect.

\textbf{KIC 3749404}: (Figure 11, Table 6)

This 20-day, eccentric ($e=0.635$), and fastly precessing ($\dot{\omega_p}=1.2^{\circ}$ yr$^{-1}$) binary has an intermediate inclination of $i=62^{\circ}$ (Hambleton et al.\ 2016). This orientation slightly favours the $m=2$ modes but $m=0$ modes cannot be excluded. Our phase measurements suggest that: the $N=19, 20, 21, 22, 26, 27$ TEOs have phases close to $|m|=2$ modes, and the $N=17, 23$ TEOs are probably $m=0$ modes. The phases of N=24 and N=21 TEOs show large deviations ($> 2 \sigma$) from the adiabatic expectations.

$\iota$ \ \textbf{Ori}: (Figure 12, Table 7)

We use the results presented in Pablo et al. (2017). The orbital and physical parameters of the binary are listed in their Table 2, and the frequencies, amplitudes, and phases of TEOs are listed in Table 3. We did not include the frequency at $f=6  f_{orb}$, since this peak is likely from the instrumental effect or an artifact from data reduction and not likely to be a real TEO frequency.

We show the pulsation phases of TEOs at $N=23, 25, 27, 33$ orbital harmonics in Figure 12. The phase measurements from the telescope pointing (Orion I) have large error bars, excluding us from making reliable mode identification. Both the red and blue filter measurements indicate $N=25$ is probably an $m=-2$ mode.
For $N=23$ and $N=33$ TEOs, the phases at both telescope pointings (Orion I and II) are not consistent. 

The pulsation amplitude modeling of TEOs presented Pablo et al.\ (2017) suggests these four TEOs are likely $m=2$ prograde modes. The phase information mentioned above is not in contradiction with this argument. Better measurements of pulsation phases are needed to make a definitive conclusion.

\textbf{KIC 8164262}: (Figure 13, Table 8)

The argument of periastron of this binary ($\omega_p=85^{\circ}$) happens to be close to $90^{\circ}$, so the theoretical phases of $|m|=2$ modes are close to the $m=0$ modes at $\phi=0.25, 0.75$. However, the measured pulsation phases do not have the tendency to cluster around the two expected phases (Figure 13). This can be explained by the following: 
Firstly, assuming the rotational modulation signature in the light curve arises from the primary star, Hambleton et al.\ (2017) derived the inclination of the primary star to be $i=35^{\circ}$. But the orbital inclination of the binary is $65^{\circ}$, so a strong spin-orbit misalignment is present in the system. Since the pulsation is from the primary star, and the pulsation axis is usually aligned with the spin axis, the observed pulsations of TEOs cannot be modeled with the simplified equation in Section 2. And $|m|=1$ modes are also observable in this case. Secondly, Fuller et al.\ (2017) showed that the TEOs in KIC 8164262 are likely in strong resonance locking, i.e., the evolution of pulsation frequencies is in concert with the evolution of the binary orbit and the stellar spin, so that the modes are always in resonance with the driving frequency from the companion. In this case, their phases can be arbitrary.  In fact, according to the modeling of Fuller et al.\ (2017), the dominant TEO at $229$ times of orbital frequency is likely an $m=1$ mode, locked in resonance. Detailed modeling of the pulsation phases for the misaligned cases is detailed in Fuller (2017), and its application to {\it Kepler} HBs is subject to a future study.

\section{Discussion}

In this paper, the pulsation phase analysis of TEOs in HBs is by no means complete. Other HBs with TEOs include KIC 4544587 (Hambleton et al. 2013), KIC 3858884 (Maceroni et al.\ 2009), etc. There is also some evidence of TEOs in $\eta$ Carina (Richardson et al. 2018) and the exoplanet-host star HAT-P-2 (de Wit et al.\ 2017). 

In order to model the TEOs in detail, fundamental parameters such as the stellar masses, radii, effective temperatures, and binary orbital parameters are required. There are already spectroscopic observations of HBs (Smullen \& Kobulnicky 2015; Dimitrov et al.\ 2017). An updated radial velocity follow-up on Shporer et al.\ (2016) is underway. The full LC+RV analysis that can yield these accurate parameters is non-trivial and has to be performed on a one-by-one basis.

We do observe some deviations from the expected adiabatic phases in some HB systems. It is difficult to determine a prior expected deviation for a particular system. Previous studies of the face-on system KOI-54 show that the pulsation phases of the TEOs with frequencies higher than $50$ times of orbital frequency generally agree with Eq.\ (3) within $1\sigma$ (O`Leary \& Burkart\ 2014, Figure 4). For the inclined systems KIC 3230227 and KIC 4142768 (Guo et al.\ 2017; Guo et al.\ 2019), the observed TEO phases also essentially agree with Eq.\ (3) within about $1 \sigma$. 
For systems in this study, we calculate the median and mean deviation from Eq.\ (3) in units of the observed one sigma errors of TEO phases. We find that the median deviations range from 0.3$\sigma$ to 2.5$\sigma$ for KIC 8719324, KIC 9016693, KIC 4248941, KIC 5034333, and KIC 3749404. Evaluating with the mean deviation, KIC 5034333 and KIC 3749404 actually show worse agreement, with deviations of 6.2$\sigma$ and 2.0$\sigma$. It is difficult to assess the result of $\iota$ Ori since different passbands show different phases. But in general, we find moderate agreement with theory, and the deviations range from about 0.3$\sigma$ to 2.0$\sigma$ given the large uncertainties of the measurements. For KIC 8164262, we find a large disagreement of 7$\sigma$. Given its spin-orbit misalignment, this large deviation is not surprising. HD 209295 shows a median deviation of 6$\sigma$.
The deviations arise from the uncertainties in the measured time of periastron passage ($T_{\rm peri}$) and the argument of periastron ($\omega_p$), as well as the systematics due to the imperfect light curve modeling. We address the two major reasons below.

The non-adiabaticity of pulsations is one of the reasons for the observed phase deviations of TEOs.
If oscillations are non-adiabatic, the imaginary part of the stellar response in term1 of eq.\ (2) will introduce a phase offset ($\Delta \phi=\phi_{nonad}- \phi_{ad}$). Quantitative calculations for each binary system would require accurate orbital and stellar parameters which are not available here. To have a general idea of how significant this phase offset could be, we performed non-adiabatic calculations with the GYRE oscillation code (Townsend \& Teilter 2013) for a stellar model from the MESA evolution code (Paxton et al. 2011, 2013, 2015) representing a typical A-type star with a mass of $M_1=2.0M_{\odot}$. The companion star is assumed to have the same mass as the primary ($M_2=M_1$). The tidal potential from the companion star is added to the momentum equation, and the forced oscillation equations are solved following Valsecchi et al.\ (2013)\footnote{A tutorial can be found for the revised version of GYRE, dubbed GYRE-FORCE (Townsend, Meng, Guo, in preparation): \href{https://bitbucket.org/rhdtownsend/gyre/wiki/TASC5:\%20Tidal\%20Forcing\%20in\%20GYRE}{https://bitbucket.org/rhdtownsend/gyre/wiki/TASC5:\%20Tidal\%20Forcing\%20in\%20GYRE}}.
We consider the $l=2,m=0$ modes here. We scanned a grid of different binary configurations with orbital periods of $(10, 20,40)$ days and orbital eccentricities of $e=(0.2,0.4,0.6,0.8)$. 
We examine the phase shift ($\Delta \phi$) introduced by the term1 in eq.\ (2) for different driving frequencies from $1\Omega_{orb}$ to $100\Omega_{orb}$, with $\Omega_{orb}$ being the orbital frequency. The phase offset is calculated as the following:
\begin{equation}
arg[term1] = arctan \left [Imag\left (term1\right ), Re\left (term1\right ) \right ]
\end{equation}
where,
\begin{equation}
term1=\left  [(2b_l-c_l)\frac{\xi_{r}(R)}{R} + +b_l \frac{\Delta F(R)}{F(R)}\right ]. 
\end{equation}
We used the relation $\Delta F/F=\Delta L/L_r-2 \xi_r/r$ to calculate the Lagrangian flux perturbations ($\Delta F/F$) (Burkart et al.\ 2012), where the radial displacement $\xi_r/r$ and the Lagrangian luminosity perturbation $\Delta L/L_r$ are direct solutions of the non-adiabatic forced stellar oscillation equations (Valsecch et al.\ 2013). The limb darkening coefficients are taken from Burkart et al.\ (2012) as $b_2=13/40$ and $c_2=39/20$ for the Kepler passband.
The general conclusion is that most of the phase shifts ($\Delta \phi$) range from 0 to 0.15, with maximum values approaching 0.25. We show the results for two representing stellar models in Figure 14. Both models have the same mass ($M=2.0M_{\odot}$) and metallicity ($Z=0.02$). Model 1 has a radiative envelope ($T_{\rm eff}=8900K$) and Model 2 has convective atmospheric properties ($T_{\rm eff}=6800K$). Note that there exist several different schemes of implementing the convective flux perturbations in the non-adiabatic calculation (e.g., the four cases in Pesnell\ 1990). It is out of the scope of this paper to compare different schemes, and we only use the default settings in GYRE (case 1).

It can be seen in Figure 14 that the non-adiabatic phase shifts for model 1 cluster around 0.06-0.07, even though the largest offsets can as large as 0.25. For model 2 with a shallow convective surface layer, the phase shifts are generally larger, with an average value of $0.08$.
Forcing harmonics ($N\Omega_{orb}$) closer to resonances with an eigen-mode of the star tend to have larger non-adiabatic phase offsets. Strong resonances with very small detuning as well as strong radiative damping of very low-frequency modes are the two cases when the phase offsets reach maximum values. This exercise indicates that the adiabatic phase relation in equation (3) is probably a reasonable assumption for stars with radiative envelopes ($T_{\rm eff} > 7000K$) when the pulsation frequencies of TEOs are not very low. For cooler stars, it has to be used with more caution. 


Spin-orbit misalignment is also one of the factors that can cause the deviation of TEO phases from the expected values in Section 2. The pulsation axis is usually aligned with the dominant symmetry axis of the star, which is the rotation axis in most cases (Lenz 2011). This is usually adopted implicitly in asteroseismlogy and also in this study. Some rare exceptions include: 1) the pulsation axis is aligned with the magnetic field in the rapidly oscillating Ap stars (Bigot \& Dziembowski 2002; Kurtz et al.\ 2011); 2) the pulsation axis is aligned with the tidal force from the companion (Hander et al. in preparation). It is usually reasonable to assume spin-orbit alignment since its timescale is much shorter than the timescale of binary orbital evolution. However, care has to be taken for some heartbeat systems since the high eccentricity may be triggered by a distant third companion.

Other than mode identifications, variations of the amplitude and phase can offer us information on the mode damping and the orbital evolution. For KOI-54, it was found that the TEO amplitudes decrease by about $2-3\%$ over three years, which cannot be explained solely by the radiative mode damping (O`Leary \& Burkart 2014). A careful observation of these TEOs can identify modes that are undergoing three or multi-mode coupling. Parent modes that surpass the mode-coupling threshold should be subject to this kind of non-linear process, and damp energy to the daughter modes. Some of the non-orbital-harmonic daughter mode frequencies share the same fractions in units of orbital frequency, e.g., KOI-54 and KIC\ 3230227. Recently, Guo (2019, submitted) found that the non-linear mode coupling in KIC\ 3230227 has probably settled to the equilibrium state. By utilizing the amplitude equations, a detailed analysis of the non-linear mode coupling can be performed (O`Leary \& Burkart 2014; Weinberg et al.\ 2013) and is highly desirable. 

This work represents an attempt to utilize pulsation phases as a mode identification method for TEOs. Detailed analysis of each system has to wait for better
characterizations of the orbital and stellar parameters. Heartbeat binary systems are excellent laboratories for the study of tides, and their potential has not been fully exploited.

 
\acknowledgments 
 
We thank the referee for the pertinent suggestions which significantly improve the quality of this paper. We thank Bill Paxton and others to maintain the MESA stellar evolution code. We thank Rich Townsend and Meng Sun for implementing tidal effect on stellar oscillations with GYRE. We are grateful to Jim Fuller for insightful discussions. We thank Eric Ford for helpful discussions.
This work is partially supported by the Polish NCN grant: 2015/18/A/ST9/00578.

 


\begin{deluxetable}{lccccc} 
\tabletypesize{\small} 
\tablewidth{0pc} 
\tablenum{1} 
\tablecaption{Model Parameters\label{tab2}} 
\tablehead{ 
\colhead{Parameter}   & 
\colhead{KIC 8719324}      &
 \colhead{-}  &
\colhead{-}     &  
\colhead{-}     
}
\startdata 
P (days) &10.2326979(300) & &\\
$T_0$ (EB catalog)              & 55003.805236& &     \\
Orbital frequency, $f_{\rm orb}$(day$^{-1}$) & 0.0977259(3)& \\
\hline
  &(LC, Kumar) & (RV)&   \\
$T_{\rm peri}$ (BJD-2,400,000)   &55004.0099(2) &- &   \\
$e$, Orbital eccentricity              &0.5998(1) &- &     \\
$\omega_p$, argument of periastron ($^{\circ}$)     & -17.123(32)&- &     \\
$i$, Orbital inclination ($^{\circ}$)  &73.54(6) & & \\
\hline
$T_{\rm eff}$ (K)                    & $7750$ & \\
$\log g$ (cgs)  & $4.5$     &      \\ 
\hline
TEOs   &   N $(=f/f_{orb})$   & Frequency, $f$ (day$^{-1}$) & Amplitude ($10^{-3}$)  & Phase ($2 \pi$)  \\                
  &   $26$   &   2.540879(5)   & 0.64472(8)   &$0.26(1)$         \\ 
  &   $ 29$   &   $2.83407(4) $   &  0.0789(6)   &$0.87(4)$          \\ 
$\phi_{m=2}$ ($2 \pi$) &   $0.34, \ \  0.84$   &     \\
$\phi_{m=-2}$ ($2 \pi$) &   $0.16,\ \  0.66$   &    \\               
$\phi_{m=0}$ ($2 \pi$) &   $0.25, \ \ 0.75$   &                 \\ 
\enddata 
\end{deluxetable} 


\begin{deluxetable}{lccccc} 
\tabletypesize{\small} 
\tablewidth{0pc} 
\tablenum{2} 
\tablecaption{Model Parameters\label{tab2}} 
\tablehead{ 
\colhead{Parameter}   & 
\colhead{KIC 9016693}      &
 \colhead{Shporer et al.\ (2016)}  &
\colhead{-}     &  
\colhead{-}     
}
\startdata 
P (days) & 26.3680271(1163) & &\\
$T_0$ (EB catalog)              & 55002.583038& &     \\
Orbital frequency, $f_{\rm orb}$(day$^{-1}$) & 0.0379247(2)& \\
\hline
   &(LC, Kumar) & (RVs, Shporer)&   \\
$T_{\rm peri}$   &55002.436(16) & 57268.91(10)&   \\
$e$, Orbital eccentricity              & 0.725(6) & 0.596(18)&     \\
$\omega_p$, argument of periastron ($^{\circ}$)     &  102.8(17)& 108.4(17) &     \\
$i$, Orbital inclination ($^{\circ}$)  &25.6(8) & & \\
\hline
$T_{\rm eff}$ (K)                    & $7262^{+201}_{327}$ & \\
$\log g$ (cgs)  & -     &      \\ 
\hline
TEOs   &   N $(=f/f_{orb})$   & Frequency (d$^{-1}$) & Amplitude ($10^{-3}$)  & Phase ($2 \pi$)  \\                
  &   $24$   &   0.910593(7)   & 0.19238(6)   &$0.275(10)$         \\ 
$\phi_{m=2}$ ($2 \pi$) &   $0.148  \ \   0.648$   &     \\
$\phi_{m=-2}$ ($2 \pi$) &  $0.352  ,\ \   0.852 $   &    \\               
$\phi_{m=0}$ ($2 \pi$) &   $0.25, \ \ 0.75$   &                 \\ 
\enddata 
\end{deluxetable} 


\begin{deluxetable}{lccccc} 
\tabletypesize{\small}  
\tablewidth{0pc} 
\tablenum{3} 
\tablecaption{Model Parameters\label{tab3}} 
\tablehead{ 
\colhead{Parameter}   & 
\colhead{KIC 4248941}      &
 \colhead{-}  &
\colhead{-}     &  
\colhead{-}     
}
\startdata 
P (days) & 8.6445976(234)& &\\
$T_0$ (EB catalog)              & 54997.105632 & &     \\
Orbital frequency, $f_{\rm orb}$(day$^{-1}$) & 0.1156792(3)& \\
\hline
   &(LC, Kumar) & (RVs)&   \\
$T_{\rm peri}$   &54997.4694(8) & -&   \\
$e$, Orbital eccentricity              & 0.423(14) & -&     \\
$\omega_p$, argument of periastron (degree)     &  -50.5(4.3)& - &     \\
$i$, Orbital inclination (degree)  & 68.3(5.5) & & \\
\hline
$T_{\rm eff}$ (K)                    & $6750$ & \\
$\log g$ (cgs)  & 4.5    &      \\ 
\hline
TEOs   &   N $(=f/f_{orb})$   & Frequency (d$^{-1}$) & Amplitude ($10^{-3}$)  & Phase ($2 \pi$)  \\                
  &   $5$   &   0.578395(1)      & 0.48790(1)    &$ 0.545(6)     $         \\ 
$\phi_{m=2}$ ($2 \pi$) &   $0.03,  \ \   0.53$   &     \\
$\phi_{m=-2}$ ($2 \pi$) &   $ 0.47  ,\ \   0.97  $   &    \\               
$\phi_{m=0}$ ($2 \pi$) &   $0.25, \ \ 0.75$   &                 \\ 

\enddata 
\end{deluxetable} 


\begin{deluxetable}{lccccc} 
\tabletypesize{\small}  
\tablewidth{0pc} 
\tablenum{4} 
\tablecaption{Model Parameters\label{tab4}} 
\tablehead{ 
\colhead{Parameter}   & 
\colhead{KIC 5034333}      &
 \colhead{-}  &
\colhead{-}     &  
\colhead{-}     
}
\startdata 
$P_{orb}$ (days) & 6.9322800(170)& &\\
$T_0$ (EB catalog)              & 54954.027612 & &     \\
Orbital frequency, $f_{\rm orb}$(day$^{-1}$) & 0.1442527(4)& \\
\hline
   &(LC, Kumar) & (RVs)&   \\
$T_{\rm peri}$ (BJD-2400000)   &54997.4694(8) & -&   \\
$e$, Orbital eccentricity              & 0.5822(9) & -&     \\
$\omega_p$, argument of periastron ($^{\circ}$)     &  278.1(3)& - &     \\
$i$, Orbital inclination ($^{\circ}$)  & 49.88(9)  & & \\
\hline
$T_{\rm eff}$ (K)                    & $9250$ & \\
$\log g$ (cgs)  & 4.5    &      \\ 
\hline
TEOs   &   N $(=f/f_{orb})$   & Frequency (d$^{-1}$) & Amplitude ($10^{-5}$)  & Phase ($2 \pi$)  \\                
  &   $18$   &   2.596525(2)    & 17.600(15)   &$ 0.580(7)    $         \\ 
  &   $13$   &   1.875299(4)    & 15.005(28)    &$ 0.677(11)     $         \\ 
  &   $20$   &   2.885043(3)    &  14.651(21)    &$  0.858(8)      $         \\ 
  &   $27$   &    3.894829(4)    & 8.778(21)   &$  0.294(11)     $         \\ 
  &   $19$   &   2.740803(5)     & 8.017(23)     &$0.258(15)     $         \\ 
  &   $66$   &   9.521462(2)      & 7.225(3)     &$ 0.948(7)     $         \\ 
  &   $4$   &    0.576985(20)        &  6.127(73)    &$ 0.76(6)     $         \\ 
  &   $12$   &   1.731012(9)       & 6.019(5)       &$ 0.239(28)    $         \\ 
$\phi_{m=2}$ ($2 \pi$) &   $0.202,  \ \   0.702$   &     \\
$\phi_{m=-2}$ ($2 \pi$) &   $ 0.298  ,\ \   0.798  $   &    \\               
$\phi_{m=0}$ ($2 \pi$) &   $0.25, \ \ 0.75$   &                 \\ 

\enddata 
\end{deluxetable} 

\clearpage




\begin{deluxetable}{lccccc} 
\tabletypesize{\small}  
\tablewidth{0pc} 
\tablenum{5} 
\tablecaption{Model Parameters\label{tab5}} 
\tablehead{ 
\colhead{Parameter}   & 
\colhead{HD209295}      &
 \colhead{Handler et al.\ (2002)}  &
\colhead{-}     &  
\colhead{-}     
}
\startdata 
P (days) & 3.10575(10) & &\\
$T_0$              & - & &     \\
Orbital frequency, $f_{\rm orb}$(day$^{-1}$) & 0.32198(1)& \\
\hline
   &(LC) & (RVs, Handler)&   \\
$T_{\rm peri}$   &- & 51771.864(14)&   \\
$e$, Orbital eccentricity              &  - & 0.352(11)&     \\
$\omega_p$, argument of periastron ($^{\circ}$)     &   -& 31.1(20) &     \\
$i$, Orbital inclination ($^{\circ}$)  &  - &  $<$ 40-45 & \\
\hline
$T_{\rm eff}$ (K)                    & $7750$ & \\
$\log g$ (cgs)  & 4.3    &      \\ 
\hline
TEOs   &   N $(=f/f_{orb})$   & Frequency (d$^{-1}$) & Amplitude ($10^{-3}mag$)  & Phase ($2 \pi$)  \\                
  &   $8$   &    2.57593(11)       &  18.3(3)   &$  0.185(2)    $         \\ 
  &   $7$   &    2.25394(11)       &  8.4(3)   &$  0.006(5)    $         \\ 
  &   $3$   &    0.96597(11)       &  7.0(3)   &$  0.891(6)    $         \\ 
  &   $5$   &    1.60996(11)       &  4.6(3)   &$  0.550(9)    $         \\ 
  &   $9$   &    2.89792(11)       &  4.5(3)   &$  0.131(9)    $         \\ 
$\phi_{m=2}$ ($2 \pi$) &   $0.077,  \ \   0.577$   &     \\
$\phi_{m=-2}$ ($2 \pi$) &   $ 0.423  ,\ \    0.923  $   &    \\               
$\phi_{m=0}$ ($2 \pi$) &   $0.25, \ \ 0.75$   &                 \\ 

\enddata 
\end{deluxetable} 

\clearpage

\begin{deluxetable}{lccccc} 
\tabletypesize{\small} 
\tablewidth{0pc} 
\tablenum{6} 
\tablecaption{Model Parameters\label{tab6}} 
\tablehead{ 
\colhead{Parameter}   & 
\colhead{KIC 3749404}      &
 \colhead{Hambleton et al.\ (2016)}  &
\colhead{-}     &  
\colhead{-}     
}
\startdata 
P (days) & 20.3063852(795) & &\\
$T_0$ (EB catalog)              & 54981.16619& &     \\
Orbital frequency, $f_{\rm orb}$(day$^{-1}$) & 0.0492456(2)& \\
\hline
   &(LC, Kumar) & (RVs+LC, Hambleton)&   \\
$T_{\rm peri}$   &54981.5723(2) &- &   \\
$e$, Orbital eccentricity              & 0.635(5) & 0.659(6)&     \\
$\omega_p$, argument of periastron ($^{\circ}$)     &  121.6(3) & 123.2(23) &     \\
$i$, Orbital inclination ($^{\circ}$)  & 37.31(7) &62(1) & \\
\hline
$T_{\rm eff}$ (K)                    & $8000(300)$ & \\
                  & $6900(300)$ & \\
$\log g$ (cgs)  & 4.10(3)     &      \\ 
  &  4.40(4)     &      \\ 
\hline
TEOs   &   N $(=f/f_{orb})$   & Frequency (d$^{-1}$) & Amplitude ($10^{-5}$)  & Phase ($2 \pi$)  \\                
  &   $21$   &   1.034138(3)   & 8.068(4)   &$0.88(1)$         \\ 
  &   $20$   &   0.984898(4)   & 6.699(9)   &$0.93(1)$         \\ 
  &   $26$   &   1.280378(6)   & 3.739(27)   &$0.067(19)$         \\
  &   $22$   &   1.083385(5)   & 4.908(29)   &$0.87(2)$         \\ 
  &   $19$   &   0.93563(1)   & 2.66(7)   &$0.92(3)$         \\ 
  &   $7$   &   0.34475(3)   & 2.1(2)   &$ 0.05(9)$         \\ 
  &   $24$   &   1.181864(7 )   & 3.47(6)   &$0.65(2)$         \\ 
  &   $23$   &   1.132624(7)   & 3.44(7)   &$ 0.71(2)$         \\ 
  &   $5$   &   0.24624(6)   & 1.21(62)    &$0.22(19)$         \\ 
  &   $17$   &    0.83716 (3)   &  0.96(36)   &$ 0.79(10) $         \\ 
  &   $27$   &   1.32963(2)     & 0.91(28)   &$0.92(8)$         \\  
$\phi_{m=2}$ ($2 \pi$) &   $0.066  \ \   0.566$   &     \\
$\phi_{m=-2}$ ($2 \pi$) &   $ 0.434  ,\ \   0.934  $   &    \\               
$\phi_{m=0}$ ($2 \pi$) &   $0.25, \ \ 0.75$   &                 \\ 
\enddata 
\end{deluxetable} 

\clearpage

\begin{deluxetable}{lccccc} 
\tabletypesize{\small} 
\tablewidth{0pc} 
\tablenum{7} 
\tablecaption{Model Parameters\label{tab7}} 
\tablehead{ 
\colhead{Parameter}   & 
\colhead{$\iota$ Ori}      &
 \colhead{-}  &
\colhead{-}     &  
\colhead{-}     
}
\startdata 

P (days) & 29.13376 & &\\
$T_{\rm peri}$ (HJD-2400000)      &51121.658(fixed) & &  \\
Orbital frequency, $f_{\rm orb}$(day$^{-1}$) & 0.034324& \\
$e$, Orbital eccentricity              & $0.7452^{+0.0010}_{-0.0014}$&  &   \\
$\omega_p$, argument of periastron (degree)      & 122.15(11) &  &    \\
$i$, Orbital inclination (degree)   &$62.86^{+0.17}_{-0.14}$ & &\\
\hline
$T_{\rm eff}$ (K)                    & $31000$ & \\
                  & $18319^{+531}_{-758}$ &  \\
\hline
TEOs   &   N $(=f/f_{orb})$   & Table 3 in Pablo et al.\ (2017)&   &   \\                
  &   $23$   &      &    &         \\ 
  &   $25$   &      &    &         \\   
  &   $27$   &      &    &        \\   
  &   $33$   &      &    &        \\   
$\phi_{m=2}$ ($2 \pi$) &   $0.071,  \ \   0.571$   &     \\
$\phi_{m=-2}$ ($2 \pi$) &   $ 0.429  ,\ \   0.929  $   &    \\               
$\phi_{m=0}$ ($2 \pi$) &   $0.25, \ \ 0.75$   &                 \\ 
\enddata 
\end{deluxetable}

\begin{deluxetable}{lccccc} 
\tabletypesize{\small}  
\tablewidth{0pc} 
\tablenum{8} 
\tablecaption{Model Parameters\label{tab8}} 
\tablehead{ 
\colhead{Parameter}   & 
\colhead{KIC 8164262}      &
 \colhead{Hambleton et al.\ (2018)}  &
\colhead{-}     &  
\colhead{-}     
}
\startdata 
P (days) & 87.4571700(6381)& &\\
$T_0$ (EB catalog)              & 54969.411534 & &     \\
Orbital frequency, $f_{\rm orb}$(day$^{-1}$) & 0.01143417(8)& \\
\hline
   &(LC, Kumar) & (LC+RVs, Hambleton)&   \\
$T_{\rm peri}$   & & 55668.829 \tablenotemark{a}&   \\
$e$, Orbital eccentricity              &  - & 0.886(3)&     \\
$\omega_p$, argument of periastron ($^\circ$)     &  -&  84.79(57) &     \\
$i$, Orbital inclination ($^\circ$)  &  - & 65(1)& \\
\hline
$T_{\rm eff}$ (K)                    & $6890(80)$ & \\
                    & $ \approx 3500$ & \\
$\log g$ (cgs)  & 3.9(1)    &      \\ 
  & -    &      \\ 
\hline
TEOs   &   N $(=f/f_{orb})$   & Frequency (d$^{-1}$) & Amplitude ($10^{-5}$)  & Phase ($2 \pi$)  \\                
  &   $229$   &   2.6184922(3)       &  1010(20)   &$  0.4526(2)     $         \\ 
  &   $241$   &   2.755 699(9)       &  35.3(8)   &$  0.723(3)    $         \\ 
  &   $123$   &   1.406 45(1)       &  22.9(9)   &$  0.601(5)    $         \\ 
  &   $158$   &   1.806 65(2)       &  15.2(8)   &$  0.331(8)    $         \\ 
  &   $124$   &   1.417 86(2)       &  15.1(9)   &$  0.852(10)    $         \\ 
  &   $132$   &   1.509 33(2)       &  13.3(9)   &$  0.840(10)    $         \\ 
  &   $194$   &    2.218 31(2)      &  12.3(8)   &$  0.196(11)    $         \\ 
  &   $128$   &    1.463 60(3)      &  11.8(9)   &$  0.419(11)    $         \\ 
  &   $317$   &    3.624 72(3)      &  9.5(8)   &$  0.384(13)    $         \\ 
  &   $129$   &    1.475 01(4)       &  8.3(9)   &$  0.889(16)    $         \\ 
  &   $125$   &    1.429 31(4)       &  6.9(9)   &$  0.920(16)    $         \\ 
  &   $137$   &    1.566 44(4)       &  6.8(8)   &$  0.366(16)    $         \\ 
  &   $114$   &    1.303 55(5)       &  6.4(8)   &$  0.761(16)    $         \\ 
  &   $264$   &    3.018 70(6)       &  5.6(8)   &$  0.507(16)    $         \\ 
  &   $22$    &    0.251 21(5)      &  5.6(8)   &$  0.729(16)    $         \\ 
$\phi_{m=2}$ ($2 \pi$) &   $0.279,  \ \   0.779$   &     \\
$\phi_{m=-2}$ ($2 \pi$) &   $  0.221  ,\ \   0.721  $   &    \\               
$\phi_{m=0}$ ($2 \pi$) &   $0.25, \ \ 0.75$   &                 \\ 
\enddata 
\tablenotetext{a}{Primary Minimum}
\end{deluxetable} 

\clearpage



\begin{figure} 
\begin{center} 
{\includegraphics[height=12cm]{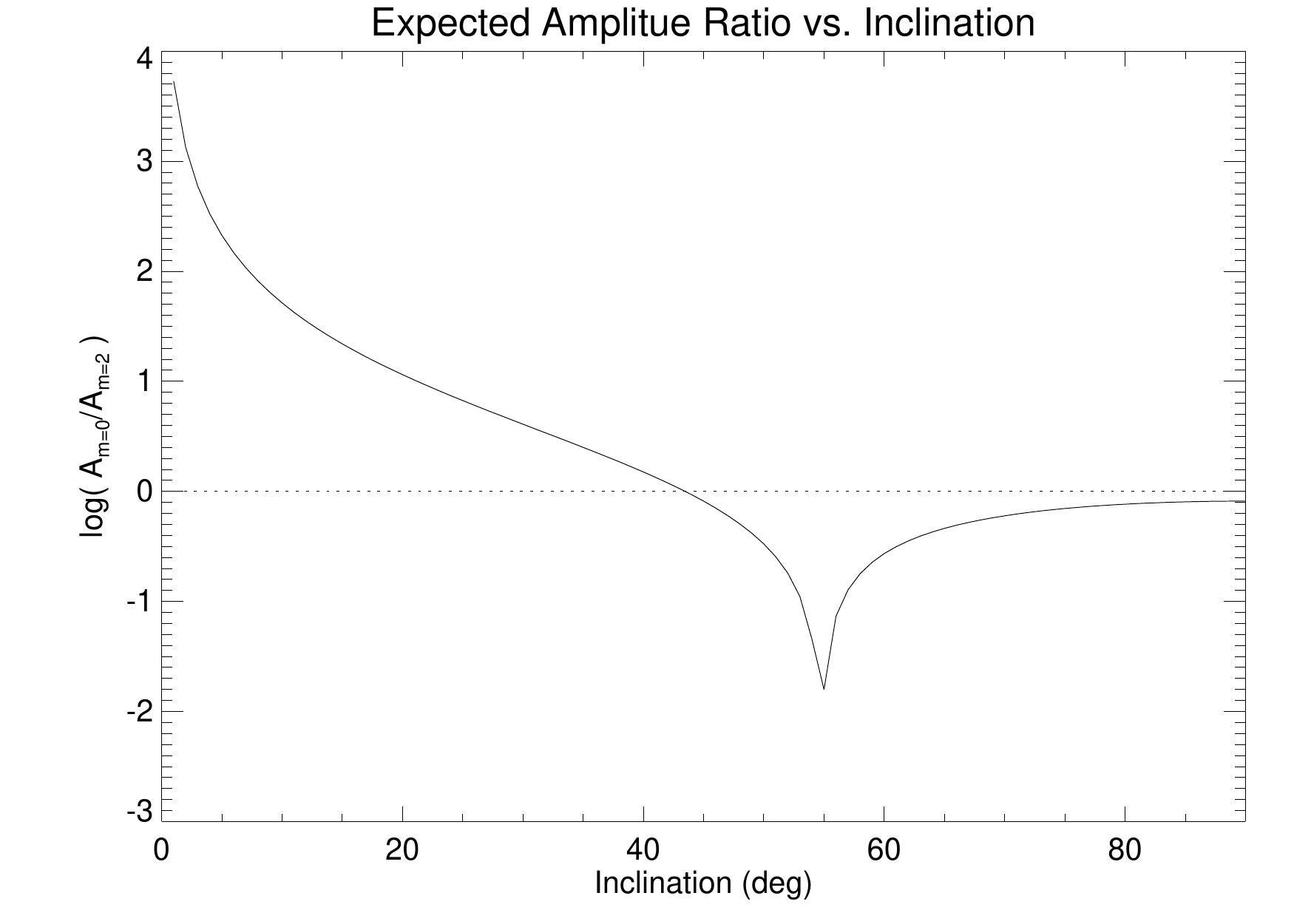}} 
\end{center} 
\caption{Amplitude ratio of $l=2, m=0$ and $l=2,m=2$ modes as a function of orbital inclination. The dotted line indicates an amplitude ratio of $1.0$.}
\end{figure}

\begin{figure} 
\begin{center} 
{\includegraphics[height=12cm]{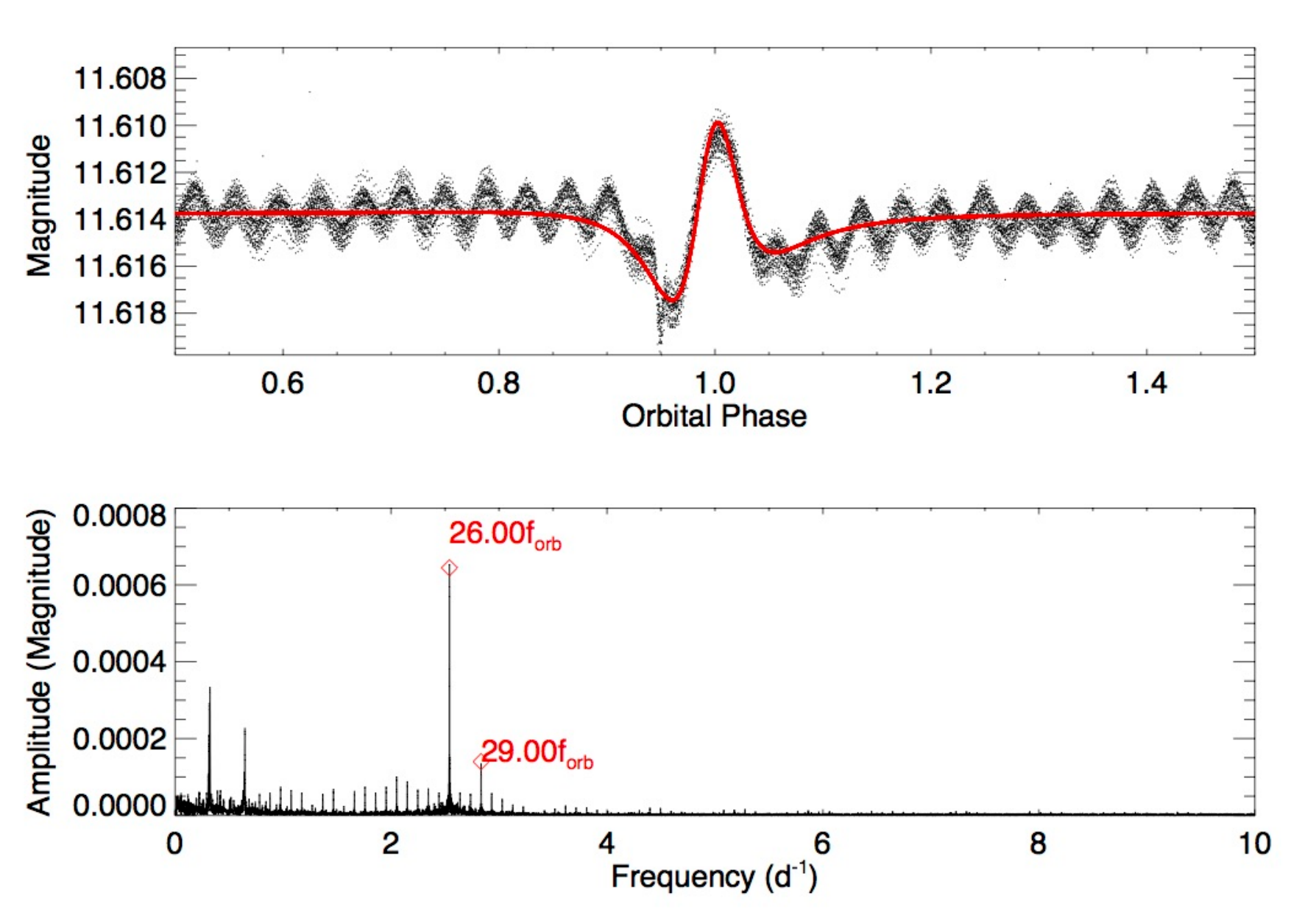}} 
\end{center} 
\caption{Phase-folded Kepler light curve of KIC 8719324 and its fourier spectrum. The two dominant TEOs at 26 and 29 times of orbital frequency ($f_{\rm orb}$) are labled.}
\end{figure}

\begin{figure} 
\begin{center} 
{\includegraphics[height=12cm]{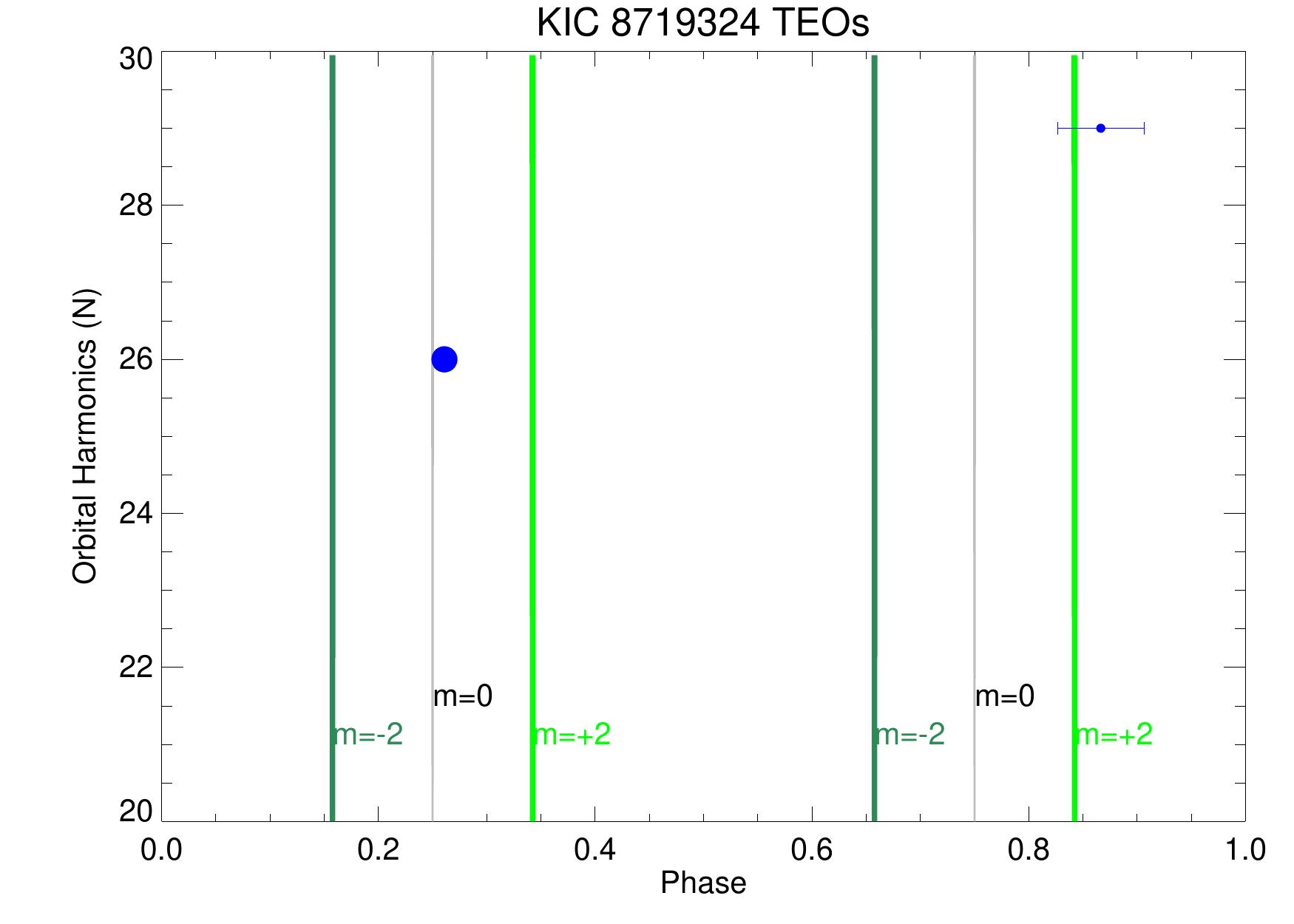}} 
\end{center} 
\caption{Pulsation Phases of the two TEOs in KIC8719324. The phases of the 26 and 29 orbital harmonics agree with the theoretical predictions for $l=2$, $m=0$ and $m=2$ modes, respectively.}
\end{figure}

\begin{figure} 
\begin{center} 
{\includegraphics[height=12cm]{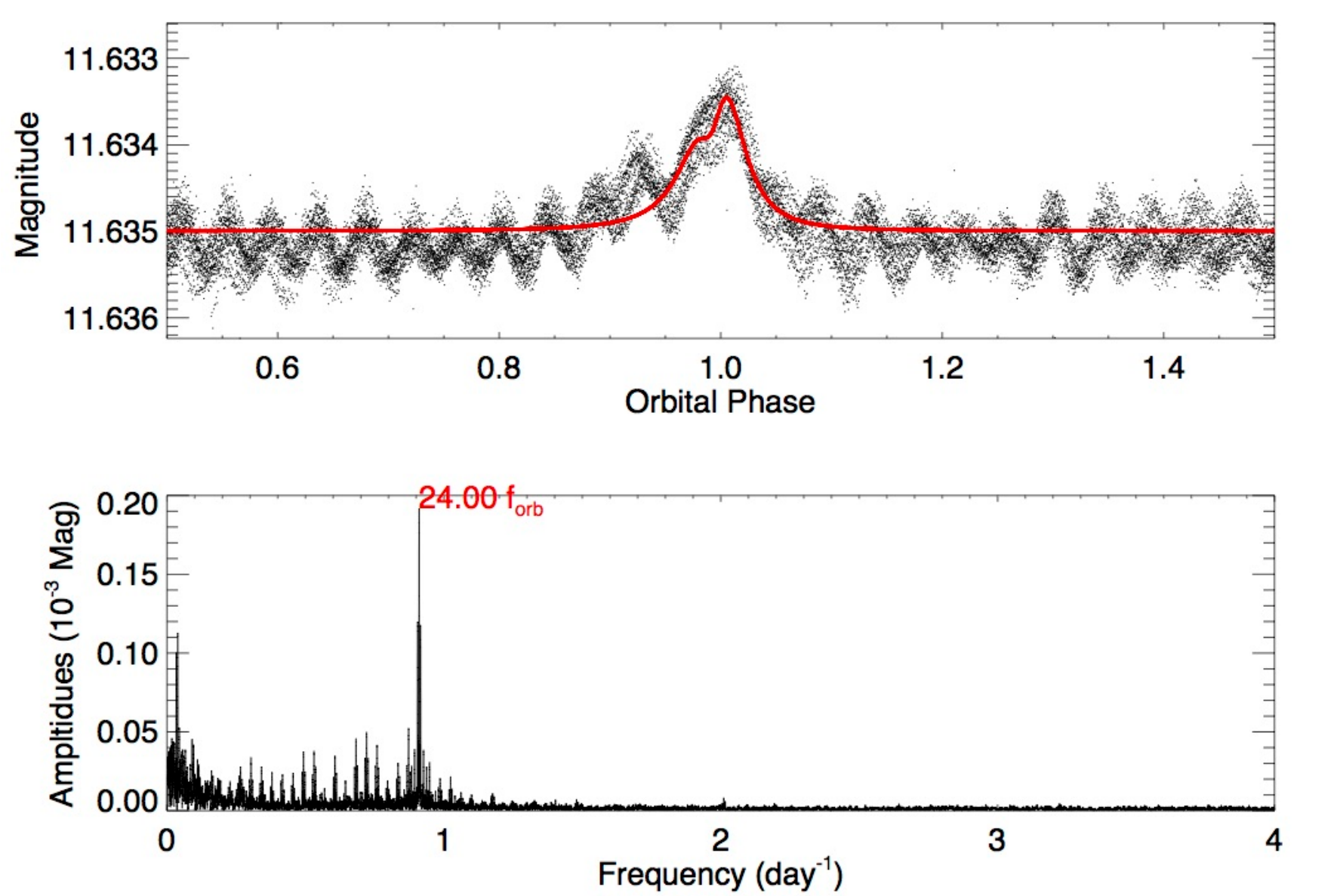}} 
\end{center} 
\caption{Phase-folded Kepler light curve of KIC 9016693 and its fourier spectrum. The  dominant TEO at 24 times of orbital frequency ($f_{\rm orb}$) is labled.}
\end{figure}


\begin{figure} 
\begin{center} 
{\includegraphics[height=12cm]{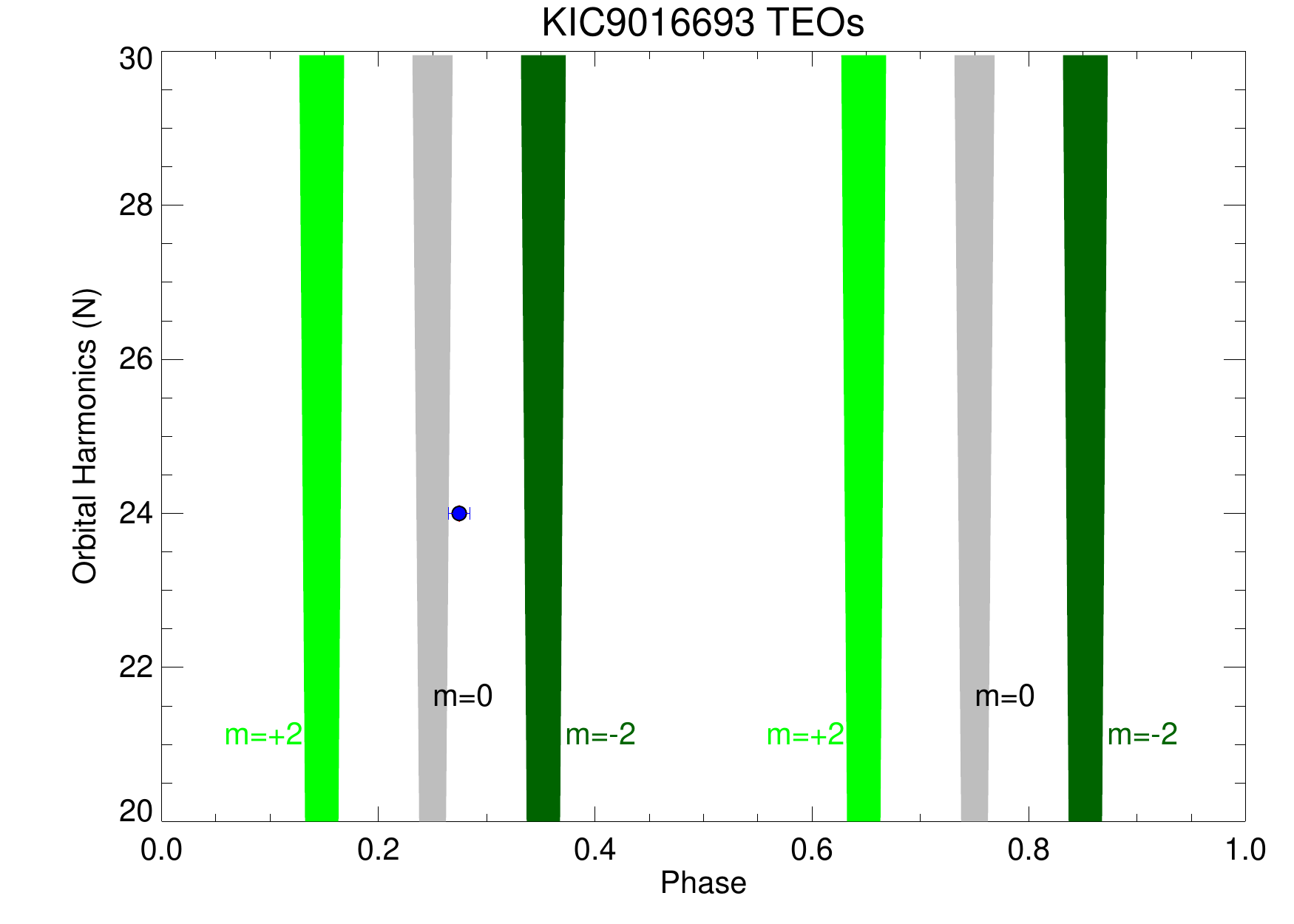}} 
\end{center} 
\caption{Pulsation Phases of the dominant TEOs in KIC 9016693 . The phase of the 24 orbital harmonics agrees with the theoretical predictions for an $l=2$, $m=0$ mode. }
\end{figure}


\begin{figure} 
\begin{center} 
{\includegraphics[height=16cm,angle=270]{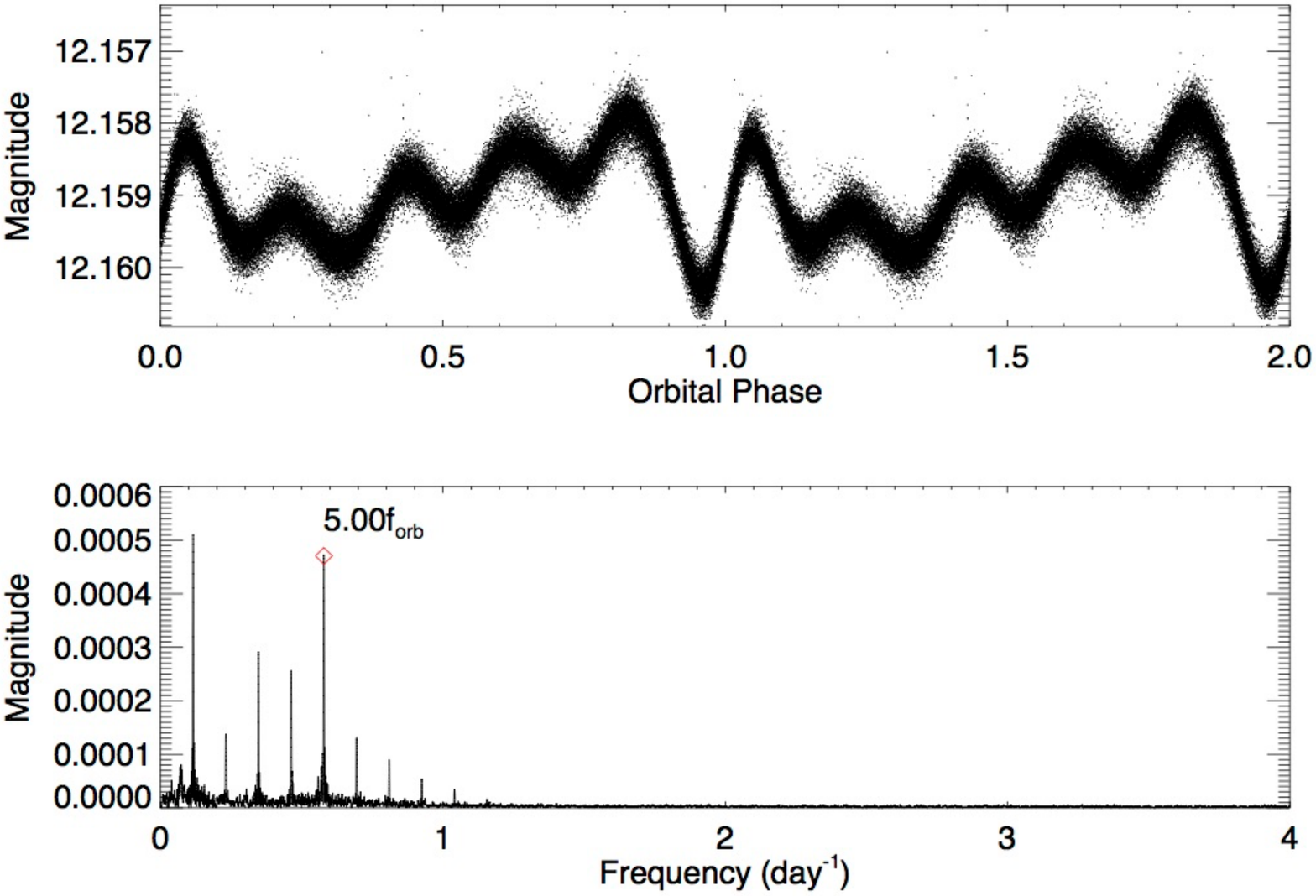}} 
\end{center} 
\caption{Phase-folded {\it Kepler} light curve of KIC 4248941 and its Fourier spectrum. }
\end{figure}

\begin{figure} 
\begin{center} 
{\includegraphics[height=12cm]{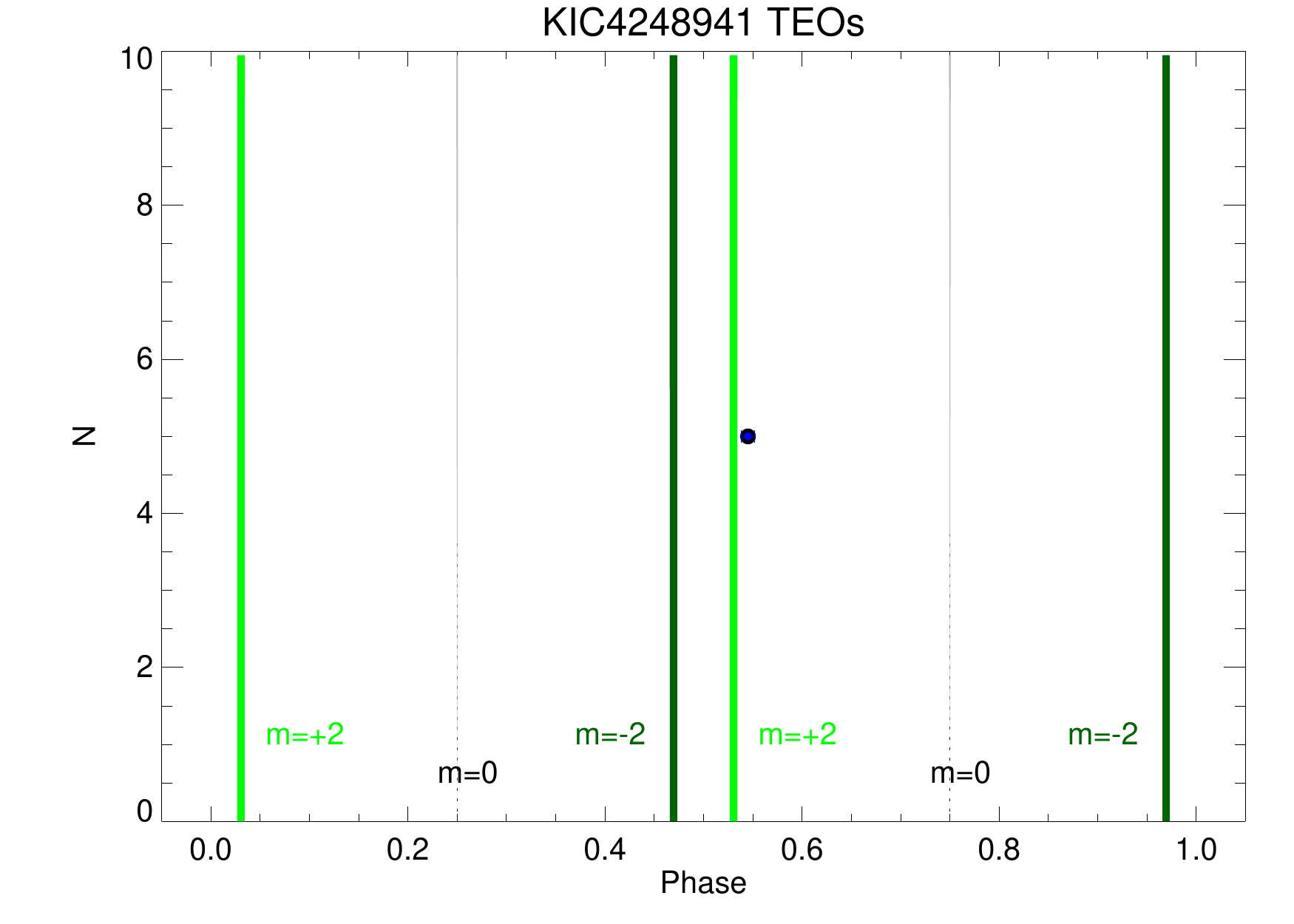}} 
\end{center} 
\caption{TEO phases of KIC 4248941. The dominant pulsation at five times of orbital frequency agrees with the $m=2$ interpretation. }
\end{figure}

\begin{figure} 
\begin{center} 
{\includegraphics[height=12cm]{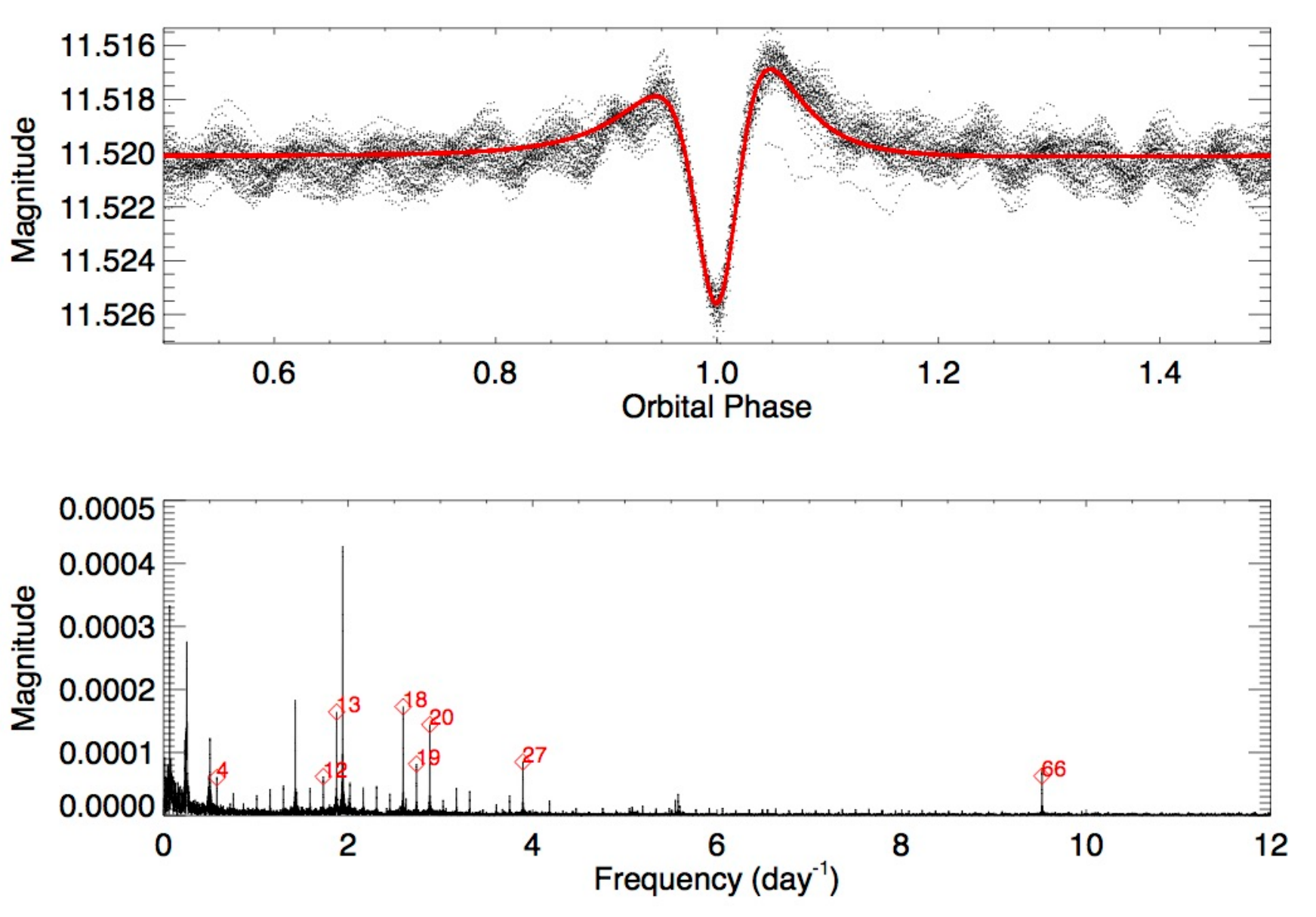}} 
\end{center} 
\caption{\textbf{Upper}: The {\it Kepler} light curve (black) and the best-fit Kumar model (red) of KIC 5034333. \textbf{Lower}: Fourier spectrum of the residual light curve. The harmonic number (N) of TEOs have been labled.}
\end{figure}

\begin{figure} 
\begin{center} 
{\includegraphics[height=12cm]{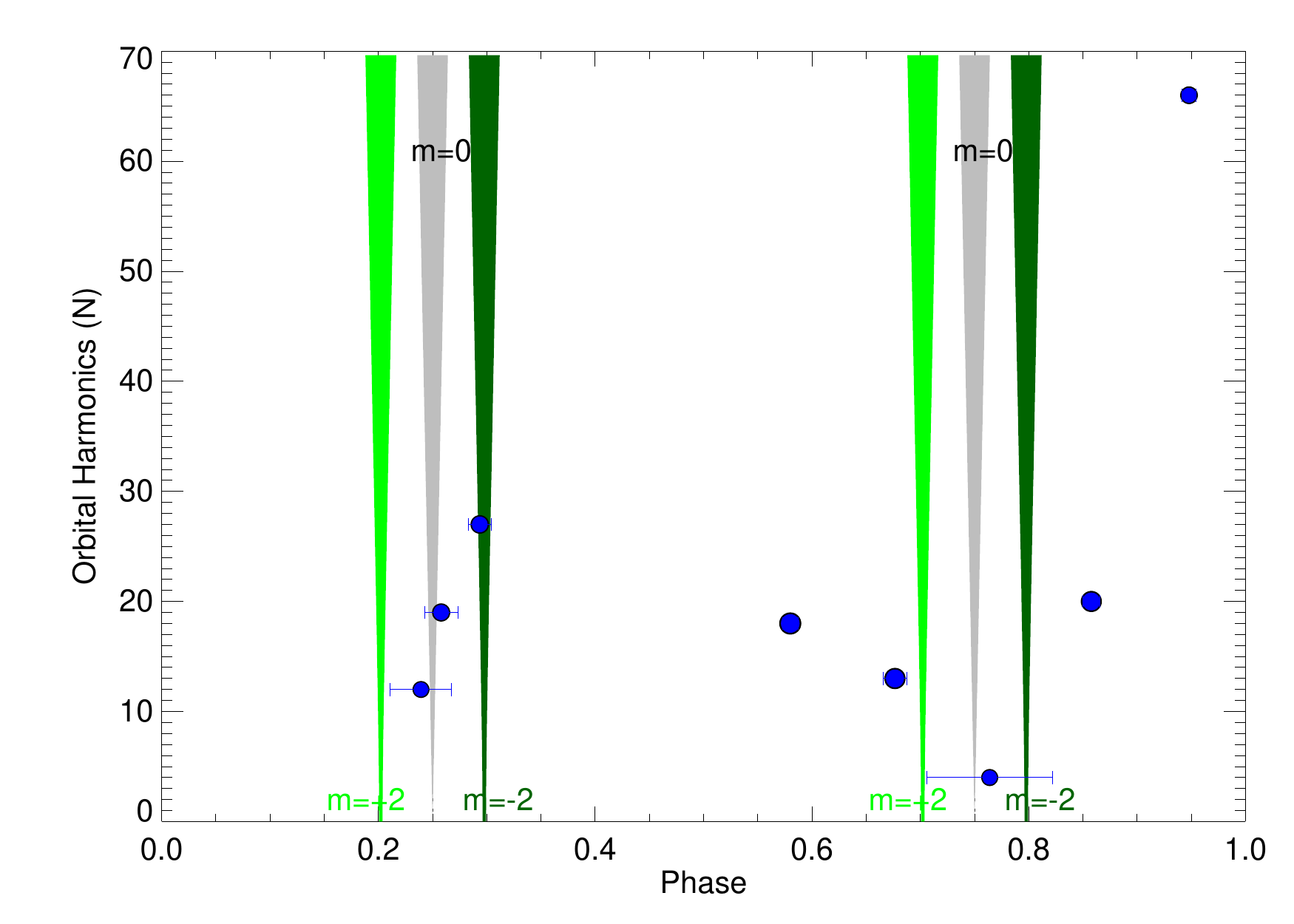}} 
\end{center} 
\caption{TEO phases of KIC 5034333 . }
\end{figure}

\begin{figure} 
\begin{center} 
{\includegraphics[height=12cm]{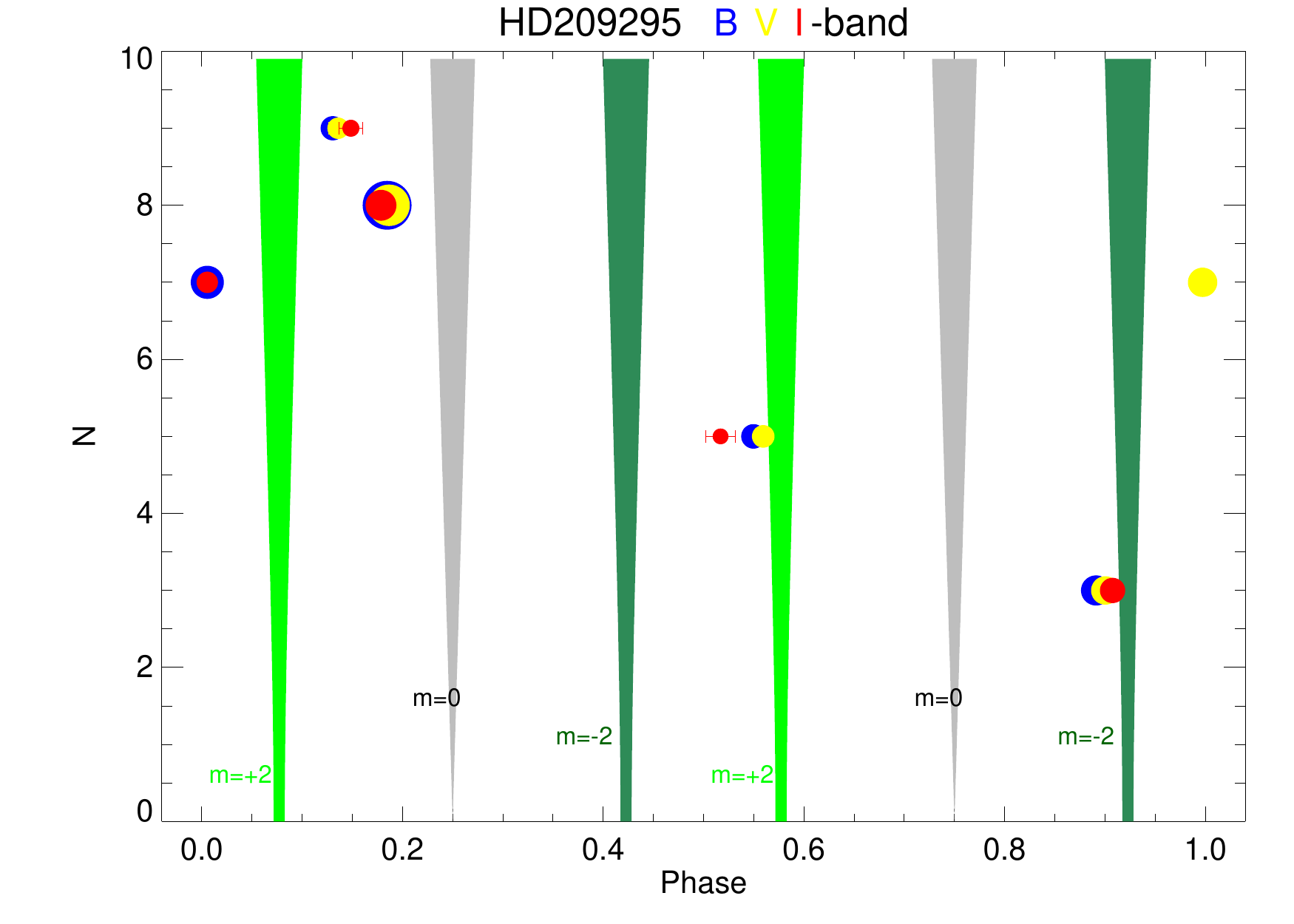}} 
\end{center} 
\caption{Pulsation Phases of the dominant TEOs in HD209295 in B (blue), V (yellow), and I-band (red), adopted from Handler et al.\ (2002)}
\end{figure} 

\begin{figure} 
\begin{center} 
{\includegraphics[angle=0,height=12cm]{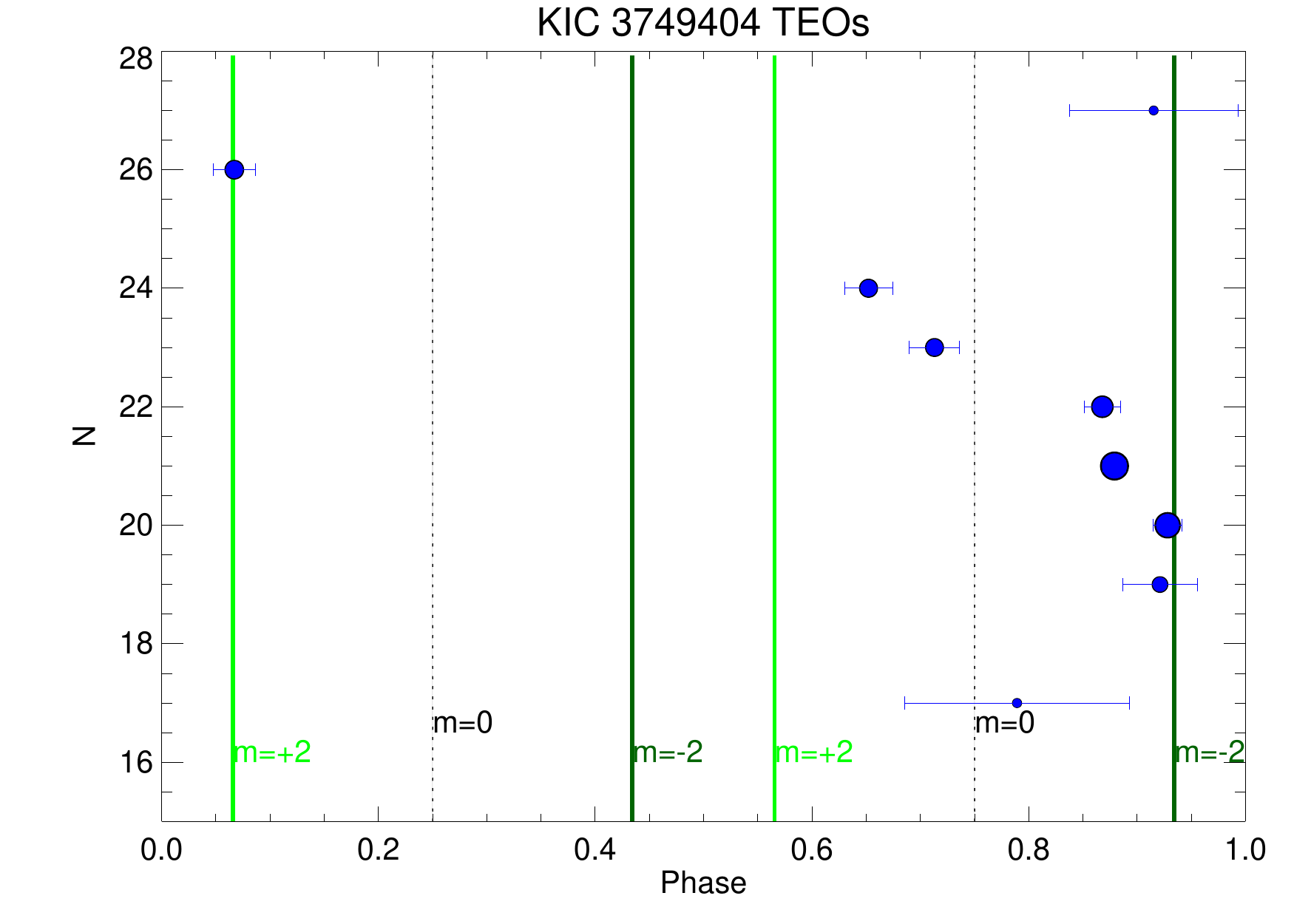}} 
\end{center} 
\caption{Pulsation Phases of the TEOs in KIC 3749404 (Hambleton et al. 2016). }
\end{figure}

\begin{figure} 
\begin{center} 
{\includegraphics[height=12cm]{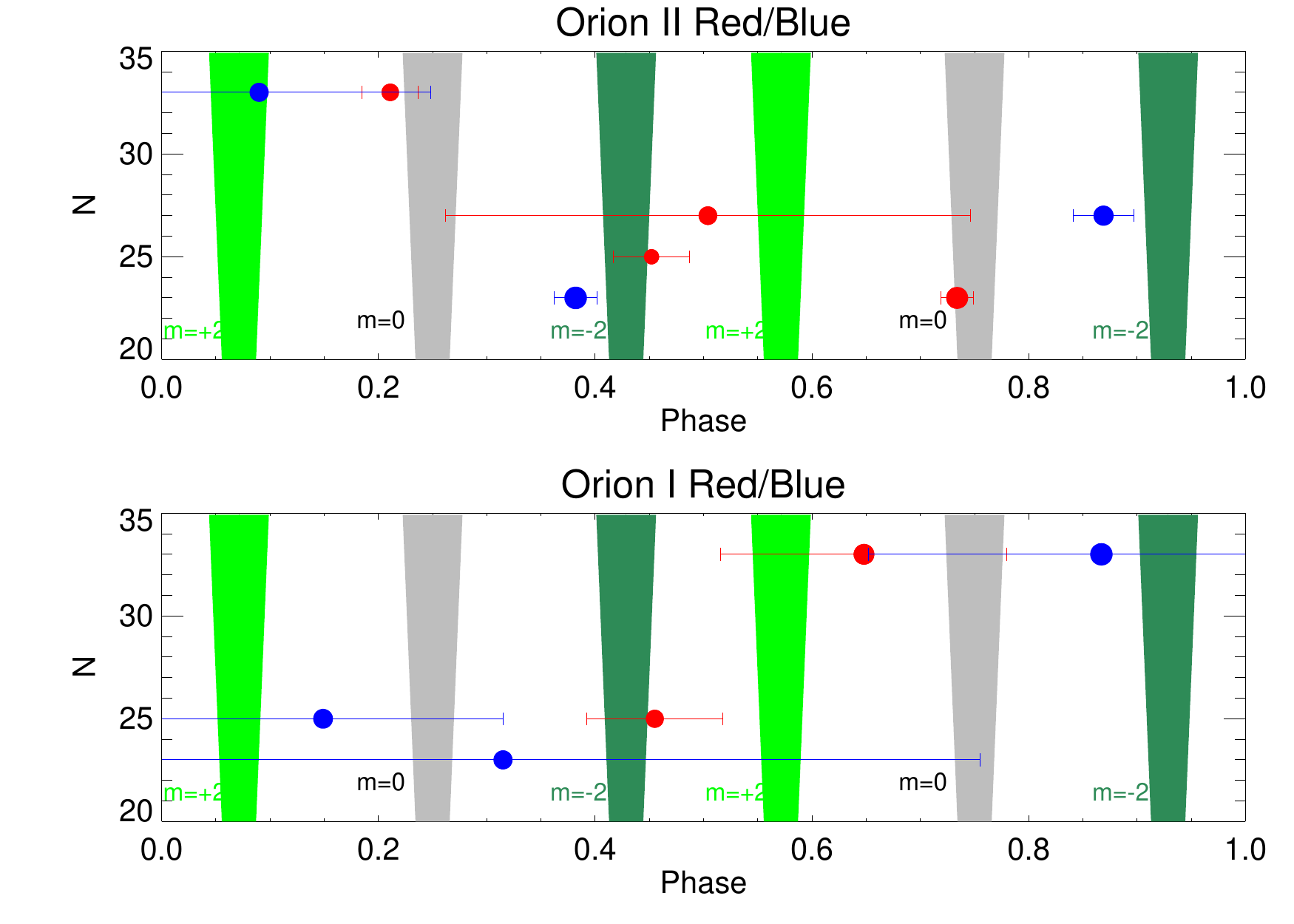}} 
\end{center} 
\caption{Pulsation Phases of the TEOs in $\iota$ Ori. The upper and lower panels show the measurements in the two telescope pointings Orion II and Orion I, respectively. Blue and red symbols indicate the measured phases in the blue and red filter of the BRITE satellite. }
\end{figure}

\begin{figure} 
\begin{center} 
{\includegraphics[height=12cm]{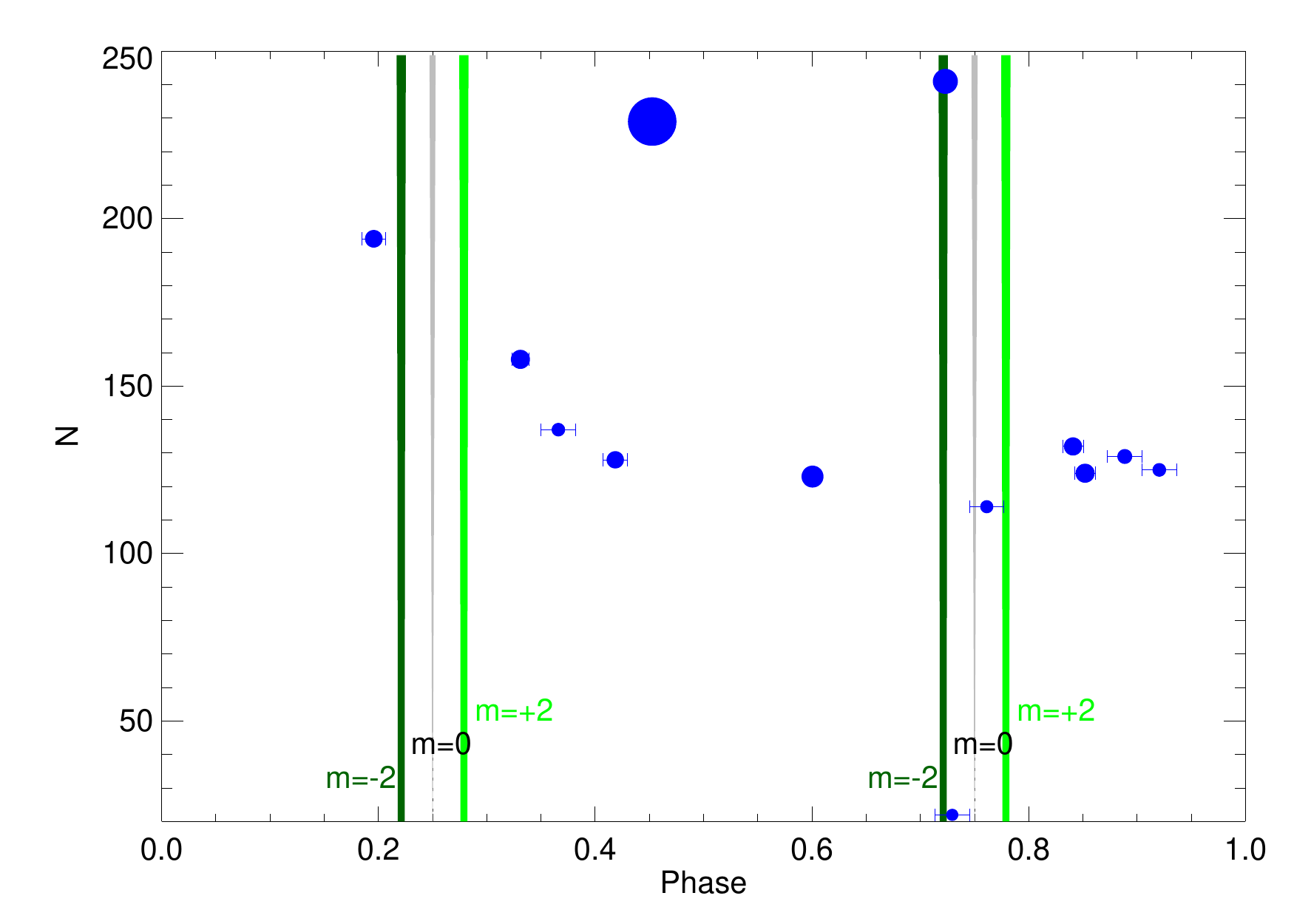}} 
\end{center} 
\caption{Pulsation Phases of the TEOs in KIC 8164262. These TEO phases do not comply with the theoretical predictions for aligned systems.}
\end{figure} 

\begin{figure} 
\begin{center} 
{\includegraphics[height=12cm]{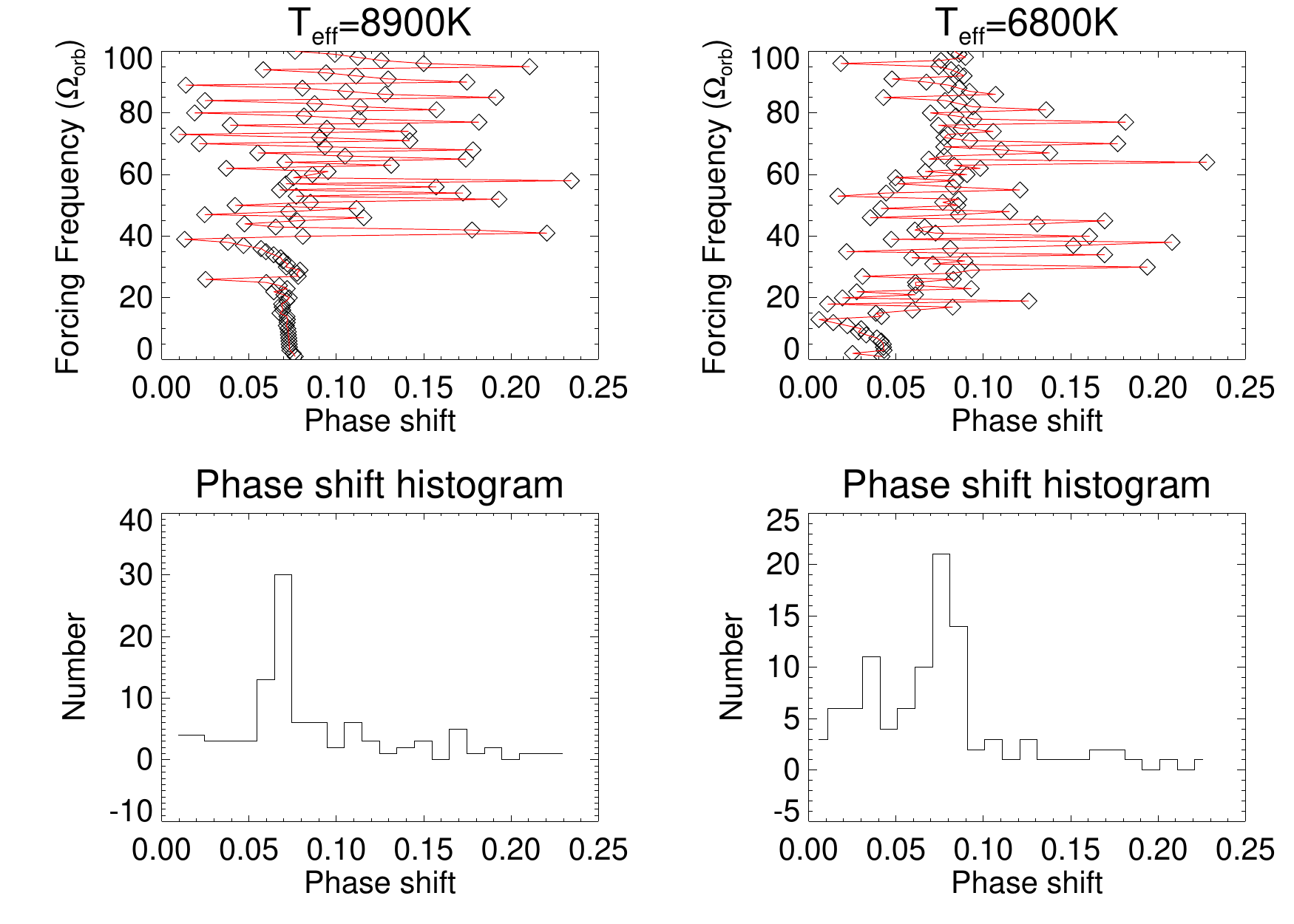}} 
\end{center} 
\caption{Phase shift arising from the Lagrangian flux perturbation $\Delta J/J$ at the stellar surface of $l=2,m=0$ modes from our nonadiabatic calculations. The upper panels show the phase shift as a function of the forcing frequencies (in units of 
the orbital frequency $\Omega_{orb}$) for two models with the same mass ($M=2.0M_{\odot}$) but different atmospheric properties ($T_{\rm eff} = 8900K$ and $T_{\rm eff}= 6800K$).}
\end{figure}

\end{document}